\documentclass[english]{article}
\usepackage{lmodern}
\usepackage{booktabs}

\usepackage[T1]{fontenc}
\usepackage[latin9]{inputenc}

\makeatletter

\usepackage{mathrsfs}   
\usepackage{slashed}     
\usepackage{bbold}  
\usepackage{url}
\usepackage{graphicx}
\usepackage[colorlinks=true,linkcolor=redLinks,citecolor=greenLinks,urlcolor=redLinks, pdfborder={0 0 1}]{hyperref}
\usepackage{color} 
\usepackage[numbers,sort&compress]{natbib}

\usepackage{amsmath}
\usepackage{amssymb}
\usepackage{breqn}
\usepackage{subcaption}
\usepackage{caption}
\usepackage{mciteplus}
\usepackage{authblk}

\definecolor{greenLinks}{rgb}{0, 0.6, 0} 
\definecolor{blueLinks}{rgb}{0, 0, 0.6}
\definecolor{redLinks}{rgb}{0.6, 0, 0}
\definecolor{eprintLinks}{rgb}{0.4, 0.4, 0.4}
\definecolor{journalLinks}{rgb}{0.6, 0, 0}

\usepackage{multirow}
\let\orig@Hy@EveryPageAnchor\Hy@EveryPageAnchor
\def\Hy@EveryPageAnchor{%
    \begingroup
    \hypersetup{pdfview=Fit}%
    \orig@Hy@EveryPageAnchor
    \endgroup
}

\let\oldFootnote\footnote
\newcommand\nextToken\relax

\renewcommand\footnote[1]{%
    \oldFootnote{#1}\futurelet\nextToken\isFootnote}

\newcommand\isFootnote{%
    \ifx\footnote\nextToken\textsuperscript{,}\fi}

\makeatother

\usepackage{babel}
\begin{document}

%
%
%
%

\title{{\Large{}\vspace{-1.0cm}} \hfill {\normalsize{}IFIC/18-20}\\*[10mm]
  {\huge{}Systematic classification of three-loop realizations of
    the Weinberg operator}{\Large{}\vspace{0.5cm}}}
\date{}

\author[1]{{\Large{}Ricardo Cepedello}\thanks{E-mail:  ricepe@ific.uv.es}}
\author[2]{{\Large{}Renato M. Fonseca}\thanks{E-mail: fonseca@ipnp.mff.cuni.cz}}
\author[1]{{\Large{}Martin Hirsch}\thanks{E-mail: mahirsch@ific.uv.es}}

\affil[1]{\small AHEP Group, Instituto de F\'isica Corpuscular, CSIC - Universitat de Val\`encia \protect\\
	Edificio de Institutos de Paterna, Apartado 22085, E--46071 Val\`encia,
	Spain}
\affil[2]{Institute of Particle and Nuclear Physics, Faculty  of  Mathematics  and  Physics, Charles  University \protect\\
	 V Hole\v{s}ovi\v{c}k\'ach 2, 18000 Prague 8, Czech Republic}

\maketitle

\begin{center}
\today
\par\end{center}
\begin{abstract}

    We study systematically the decomposition of the Weinberg operator
    at three-loop order. There are more than four thousand connected
    topologies. However, the vast majority of these are infinite
    corrections to lower order neutrino mass diagrams and only a very
    small percentage yields models for which the three-loop diagrams
    are the leading order contribution to the neutrino mass matrix. We
    identify 73 topologies that can lead to genuine three-loop models
    with fermions and scalars, i.e. models for which lower order
    diagrams are automatically absent without the need to invoke
    additional symmetries. The 73 genuine topologies can be divided
    into two sub-classes: Normal genuine ones (44 cases) and special
    genuine topologies (29 cases). The latter are a special class of
    topologies, which can lead to genuine diagrams only for very
    specific choices of fields.  The
    genuine topologies generate 374 diagrams in the weak basis, which
    can be reduced to only 30 distinct diagrams in the mass eigenstate
    basis. We also discuss how all the mass eigenstate diagrams can be
    described in terms of only five master integrals. We present 
    some concrete models and for two of them we give numerical
    estimates for the typical size of neutrino masses they
    generate. Our results can be readily applied to construct other
    $d=5$ neutrino mass models with three loops. \\ \\ \\

\end{abstract}
\newpage{}

\section{Introduction}

The smallness of the observed neutrino masses has motivated many
theoretical studies. For Majorana neutrinos, it
 can be understood from the Weinberg operator
\cite{Weinberg:1979sa}:
\begin{equation}\label{eq:OW}
  {\cal O}_W = \frac{c_{\alpha\beta}}{\Lambda} L_\alpha L_\beta H H
\end{equation}
The classical seesaw picture of this operator
\cite{Minkowski:1977sc,Yanagida:1979as,Mohapatra:1979ia,Schechter:1980gr}
corresponds to choosing for $\Lambda$ a very large scale, say $\Lambda
\sim {\cal O}(10^{14})$ GeV, in which case neutrino masses are of the
(sub-)eV order for $c_{\alpha\beta} \simeq {\cal O}(1)$.  However,
$c_{\alpha\beta}$ could be naturally much smaller than one, resulting
in correspondingly lower values for the energy scale $\Lambda$ at
which lepton number is violated. There are two simple ways of
realizing such a suppression: (i) neutrino masses might be radiatively
generated, in which case $c_{\alpha\beta} \propto 1/(16 \pi^2)^n$,
where $n$ is the number of loops; (ii) higher d-dimensional operators
might be responsible for neutrino mass generation. Note that such
operators are always of the form ${\cal O}_W \times
(H^{\dagger}H)^{\frac{d-5}{2}}$. In this paper we will follow the
former idea and study systematically the decomposition of
eq. (\ref{eq:OW}) at three-loop order. For a recent systematic study
of higher dimensional tree-level neutrino mass models see
\cite{Anamiati:2018cuq}.

The idea that neutrino masses might be small due to their radiative
origin is nearly as old as the seesaw mechanism itself, the Zee model
being the classical example \cite{Zee:1980ai}.  Many references can be
found in the recent review \cite{Cai:2017jrq}.  See also
\cite{Farzan:2012ev} for general recipes on building loop neutrino
mass models. For our work, the most relevant references in the
literature are \cite{Ma:1998dn,Bonnet:2012kz,Sierra:2014rxa}: In
\cite{Ma:1998dn} it was pointed out that there are just three variants
of the seesaw at tree-level; references \cite{Bonnet:2012kz} and
\cite{Sierra:2014rxa} gave a systematic decomposition of the Weinberg
operator at 1-loop and 2-loop level, respectively. We make use of
these results in the current work, which can be understood as a
extension of \cite{Bonnet:2012kz,Sierra:2014rxa} to 3-loop diagrams.
We also mention in passing that $d=7$ neutrino masses were studied
systematically at tree-level in \cite{Bonnet:2009ej} and at the 1-loop
level in \cite{Cepedello:2017eqf}. The only genuine $d=7$ tree-level
model was discussed in \cite{Babu:2009aq}.

All the above papers follow a diagramatic approach to the
classification of neutrino mass models. Alternatively, one can start
from a list of non-renormalizable $\Delta L=2$ operators and find the
ultra-violet completions (i.e. neutrino mass models) from a
deconstruction of all operators. This approach has been used, for
example, in \cite{Babu:2001ex,deGouvea:2007qla,Angel:2012ug},
see also the discussion in \cite{Cai:2017jrq}.

Although understandably 3-loop neutrino mass models have received much
less attention than lower order ones, still a number of papers on the
subject can be found in the literature. An early example where the
Weinberg operator is generated via a 3-loop diagram is the so-called
KNT model \cite{Krauss:2002px}. In it, two charged scalar singlets
plus a right-handed neutrino are added to the Standard Model
(SM). Since the main motivation of this paper is to connect neutrino
masses with dark matter, a $Z_2$ symmetry is added by hand under which
one scalar and the right-handed neutrino are odd. The resulting 3-loop
diagram is shown as model-2 in section \ref{sec:examples} and we label
its topology as $T_5$ in appendix \ref{sec:genuinetopologies}. The KNT
model and a number of variants based on this original idea have been
studied in several papers since then. For example, the authors of
\cite{Cheung:2004xm} added a second $N_R$ to explain the two distinct
mass scales observed in neutrino oscillations. Other variants using
$SU(2)$ triplets and quintuplets instead of singlets were discussed in
\cite{Ahriche:2014cda} and in \cite{Ahriche:2014oda},
respectively. Similar ideas revolving around the use of larger
representations in the KNT model have been discussed in
\cite{Chen:2014ska}. Also, the use of coloured particles in the KNT
loop was studied in \cite{Gu:2012tn,Nomura:2016ezz,Cheung:2016frv}.
Yet another model variation with doubly charged singlets and doubly
charged vector-like fermions was constructed in
\cite{Okada:2016rav}. A scale-invariant version of the KNT model was
presented in \cite{Ahriche:2015loa}.  Finally, we mention that the
phenomenology associated to the scalars in the KNT model and its
triplet variant was discussed in \cite{Ahriche:2015taa}; for the
collider phenomenology see \cite{Ahriche:2015lqa,Ahriche:2014xra}.  In
all the above papers on the KNT model, the $Z_2$ symmetry was
introduced by hand.  In \cite{Ahriche:2015wha}, however, $SU(2)_L$
septuplets are used (both scalar and fermionic), which lead to an
accidental $Z_2$ symmetry and thus to automatically stable dark
matter. There is also a very recent paper \cite{Hati:2018fzc},
based on the KNT topology, which uses (triplet) leptoquarks
to explain also the anomalies observed in B-decays.

There are other 3-loop models based on topologies different from the
one of the KNT model. The AKS model \cite{Aoki:2008av} requires two
Higgs doublets, two $N_R$, a charged and a neutral scalar singlet.  A
$Z_2$, under which singlets are odd, eliminates the tree-level seesaw
and stabilizes again the dark matter candidate. The AKS model generates
two neutrino mass diagrams: They correspond to our diagrams $D_6^M$
and $D_7^M$ in figure \ref{fig:normalgenuinediagrams}, descending from
topologies $T_{37}$ and $T_{22}$, respectively, shown in appendix
\ref{sec:genuinetopologies}.  The phenomenology and vacuum stability
constraints for the AKS model have been studied in
\cite{Aoki:2009vf,Aoki:2010aq,Aoki:2011zg}.

The same topologies and diagrams as in the AKS model appear also in
\cite{Okada:2015hia}. However, the authors of \cite{Okada:2015hia} use
doubly charged vector-like fermions and a scalar doublet with
hypercharge $Y=3/2$ (plus the singlets of the AKS model).  The
diagrams $D_6^M$ and $D_7^M$ appear also in 
\cite{Ko:2016sxg}. Here, however, these diagrams descend from our
topologies $T_{40}$ and $T_{33}$.  $D_6^M$ appears also in a model
based on singlets \cite{Gu:2016xno}, again descending from
$T_{40}$. The last 3-loop model we mention is the one discussed in
\cite{Culjak:2015qja}. It contains a scalar septet and a
fermionic quintuplet, generating the diagrams $D_6^M$ (from $T_{37}$) and
$D_7^M$ (from $T_{32}$).  The model contains an accidental
$Z_2$, but still one needs to impose an additional $Z_2$ by hand.

Our classification scheme for the different topologies concentrates on
identifying 3-loop \textit{genuine} topologies, which are those
associated with the dominant contributions to the neutrino mass
matrix.  We discuss thoroughly the concept of ``genuineness'' in section
\ref{sect:class}. There, we also define two classes of such
topologies: Normal (or ordinary) genuine topologies (in total there
are 44 of them) and special genuine topologies, which require very
special fields (29 cases).  The full list is given in appendix
\ref{sec:genuinetopologies}.

As we will explain later, these special genuine topologies are
associated to finite loop integrals, even though they generate some
particular 3- or 4-point interaction at loop level. This happens because the
corresponding tree-level renormalizable vertex vanishes due to the
antisymmetric nature of some $SU(2)_L$ contractions. Our 29 special
genuine topologies are of this type. However, there are some 3-loop
models in the literature which also rely on such a loop-generated
vertex, but do not (necessarily) fall into that list of 29 special
topologies. Instead, those models use a symmetry to forbid the
tree-level vertex, which is then generated at loop level, once the
symmetry is (spontaneously) broken.  The model presented in
\cite{Hatanaka:2014tba} falls into this class. (It generates diagram
$D_4^M$, at the level of diagrams in the mass eigenstate basis.)
Here, the tree-level vertex $e_Re_Rk^{++}$ of the Babu-Zee 2-loop
model \cite{Cheng:1980qt,Zee:1985id,Babu:1988ki} is forbidden by a
global $U(1)$. Spontaneous breaking of this $U(1)$ to a $Z_2$
generates a Majorana mass for the $N_R$ (and produces a singlet
Majoron) and generates this vertex at the 1-loop level.  Similarly, a
3-loop diagram ($D_5^M$) appears in \cite{Nishiwaki:2015iqa}. Note
that, this type of models falls into the class called ``fermionic
cocktail'' in \cite{Cai:2017jrq}.  We also mention
\cite{Dutta:2018qei}, which studies a neutrino mass model based on an
additional $SU(2)$ group, which leads to a 3-loop diagram with new
vector bosons.

Then, there are also higher-dimensional 3-loop models.  The authors of
\cite{Cheung:2017efc} presented a 3-loop model with neutrino masses at
$d=7$. The so-called ``cocktail model'' \cite{Gustafsson:2012vj} is
actually a 3-loop model at $d=9$. (For a study of the phenomenology of
the cocktail model see \cite{Gustafsson:2014vpa}). Note that we limit
our deconstruction of 3-loop models to $d=5$ models. Thus, although
the five master integrals defined in the appendix cover all possible
3-loop integrals, our topologies (and diagrams) are complete only for
the $d=5$ case.

The rest of this paper is organized as follows. In section
\ref{sect:class} we explain our classification scheme and how our
results are obtained. All models which we classify as \textit{genuine}
have finite 3-loop integrals and thus do not need lower order counter
terms for renormalization. We discuss that there is a further class of
genuine topologies with finite 3-loop integrals, which correspond to
loop generation of some 3- or 4-point vertices.  We call these the
\textit{special genuine} topologies. We then show that the 73 genuine
topologies (normal plus special ones) are associated to 374 diagrams
in the weak basis, which get reduced to only 30 diagrams in the mass
basis. We will discuss the dichotomy between normal and special
topologies/diagrams in detail.  We would like to point out that we are
mostly interested in diagrams with new scalars (and/or new
fermions). However, there could also be diagrams with vector
particles, either the standard model W-boson or some exotic
vector. While many of our results apply also to diagrams with vectors,
we stress that due to a particular loophole in our procedure for
finding genuine diagrams our list of genuine diagrams is incomplete
for vectors. This is discussed in section 2 in more detail.  In
section \ref{sec:examples} we show some example models, discussing
them briefly. For two of them we perform also numerical calculations
of the expected neutrino mass scale; they can easily reproduce the
observed neutrino masses.  We then close with a short discussion of
our results. Several more technical aspects of our work are deferred
to the appendix, where we show the genuine topologies and give the
full definitions of the master integrals. A full list of the
topologies and diagrams mentioned in our work can be found in
\cite{extraData}.

\section{Genuine topologies, diagrams and models}\label{sect:class}

The following nomenclature will be used throughout the text. We shall
call \textit{topologies} to those Feynman diagrams where no property
of the fields is considered (in graph theory, these are also known as
undirected multigraphs). If scalars are differentiated from fermions,
we will call them \textit{diagrams}. If additionally the quantum
numbers of the internal particles are specified, we will use the
expression \textit{model-diagrams} (or just \textit{model} when it is
clear from the context what we mean by this word).

Let us discuss now the concept of a \textit{genuine} model-diagram,
diagram and topology.  Essentially, we want to identify this concept
with those model-diagrams (plus their associated diagrams and
topologies) for which the leading contribution to neutrino masses
arises at 3-loops, without the need to introduce extra symmetries.
Nevertheless, note that models with extra symmetries can be
phenomenologically interesting; we provide one example of a 3-loop
model-diagram with extra symmetries in section \ref{sec:examples}.

It is important to keep in mind that, in general, loop integrals have
a finite and an infinite part.  Infinite integrals require a lower
order counter term in a consistent renormalization scheme.  Thus,
models with infinite $n$-loop amplitudes must necessarily also
generate neutrino mass diagrams with less loops, and for that reason
model-diagrams with an infinite amplitude are not genuine in our
sense. However, certain diagrams lead automatically only to finite
loop integrals, in which case it might be possible to build genuine
model-diagrams from them.

Finiteness of the amplitude of a model-diagram is therefore a
necessary condition for it to be genuine.  However, it is not
sufficient: it is also necessary to ensure that there are no other
automatically generated model-diagrams with less loops.  Consider
neutrino masses generated via the dimension $d=5+2n$ operator
\begin{equation}
	LLHH\left(H^{\dagger}H\right)^{n}
\end{equation}
through a diagram with $\ell$ loops. It is expected that

\begin{equation}
  M_{\nu}\sim\frac{1}{\left(4\pi\right)^{2\ell}}
  \left(\frac{\left\langle H\right\rangle}{\Lambda}\right)^{1+2n}
  \left\langle H\right\rangle\,.
\end{equation}
\\ As such, for diagrams with a characteristic scale $\Lambda\sim$
TeV, removing a loop ($\ell \to \ell-1$) and increasing the operator
dimension by two units ($n \to n+1$) leaves $M_\nu$ with roughly the
same value. So in order to have a dominant $\left(d,\ell\right)$
contribution to neutrinos masses, those with
$\left(d^\prime,\ell^\prime\right)=\left(d-2i,\ell-j\right)$ and $j>i$
should be absent.\footnote{Note that by closing some pairs of
  $H$/$H^*$ external lines it will always be possible to find other
  contributions with
  $\left(d^\prime,\ell^\prime\right)=\left(d-2i,\ell-i\right)$}
\textit{Genuine} model-diagrams are those associated to these cases;
in other words, the combination of fields participating in genuine
model-diagrams must not be sufficient, by itself, to generate other
more important neutrino mass contributions.  For example, a model with
a right-handed neutrino, $\nu_R$, will also give a $d=5$ tree-level
contribution to the neutrino mass (unless an additional symmetry is
used to eliminate some unwanted couplings), which will likely be the
more important one.  Stated in this way, genuineness is a concept
which applies to model-diagrams.  However, the list of such cases is
infinite, in principle, as there are endless possibilities of
assigning quantum numbers to the internal particles. We will therefore
be interested in cataloguing those diagrams and topologies for which
there exists at least one genuine model-diagram. These topologies and
diagrams will be considered genuine themselves.

We found all such diagrams after a series of steps. First, using known
algorithms (see for example \cite{Read:1981}) we generate a list of
all 3-loop connected topologies with 3- and 4-point vertices, and four
external lines. This list contains a total of 4367 topologies.  Only
3269 of them can accommodate 2 external fermion lines plus 2 external
scalars using renormalizable interactions only.

Still at the level of topologies, we can already exclude a large
number of these by applying the following straightforward criteria.
We eliminate all cases with tadpoles (i.e., self-connecting vertices)
and self-energies (i.e., 2-point subdiagrams with one or more loops),
since these have always infinite parts in their loop integrals. This
cut leaves us with 370 topologies.

\begin{figure}[tbph]
\begin{centering}
\includegraphics[scale=0.90]{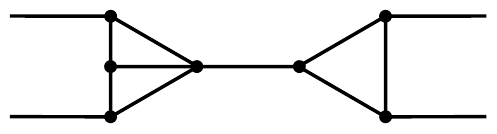}
\par\end{centering} \protect\caption{\label{fig:discon}A 1-particle
  reducible 3-loop topology. The line in the middle must correspond to
  one of the standard tree-level seesaw mediators, hence the topology
  is not genuine. The middle line is a fermion with the quantum
  numbers of $\nu_R\equiv(\boldsymbol{1},\boldsymbol{1},0)$ or
  $\Sigma\equiv(\boldsymbol{1},\boldsymbol{3},0)$ if it splits the
  external fields as $LH|LH$, or it is the scalar
  $\Delta=(\boldsymbol{1},\boldsymbol{3},-1)$ if the splitting is
  $LL|HH$.}
\end{figure}

We then eliminate the 1-particle reducible topologies, that is those
topologies which become disconnected by cutting one of its
lines. These cases can be discarded because the line which would
disconnect the topology must have the quantum numbers to mediate type
I, type II or type III seesaw (see fig. (\ref{fig:discon})).

This leaves us with 160 potentially genuine topologies, which can be
divided in three classes: \textit{normal genuine} topologies,
\textit{special genuine} topologies and \textit{non-genuine}
topologies.  This division is done in steps.

There are those topologies for which an internal loop (or loops) can
be compressed to a 3-point vertex (of the type fermion-fermion-scalar
or scalar-scalar-scalar).  For the remaining topologies, we find all
valid diagrams, labelling internal lines as scalars or fermions in all
possible ways, keeping externally exactly two scalars and two
fermions. In this list we identify all diagrams with internal loops
which can be compressed to a 4-scalar interaction.  All diagrams
without 3-point nor scalar 4-point loop subdiagrams fall into one of
44 topologies --- see the counting on table
\ref{tab:normalGenuineCounting}. These we consider \textit{normal
  genuine} diagrams and topologies. Their complete list is given in
appendix \ref{sec:genuinetopologies} (the topologies) and in
\cite{extraData} (the topologies and the diagrams).

\begin{table}
	\begin{centering}
		\begin{tabular}{cc}
			\toprule 
			All topologies (connected, with 3 loops and 4 legs) & 4367\tabularnewline
			Allow two external fermion lines & 3269\tabularnewline
			No tadpoles & 1056\tabularnewline
			No self-energies & 370\tabularnewline
			1-particle irreducible & 160\tabularnewline
			No 3-point loop subgraphs & 70\tabularnewline
			No unavoidable 4-point scalar loop subgraphs & 44\tabularnewline
			\bottomrule
		\end{tabular}
		\par\end{centering}
	
\protect\caption{\label{tab:normalGenuineCounting}Number of
  topologies, after the cumulative application of a series of cuts
  described in the text. Out of the initial 4367, only 44 have all the
  properties listed above: these are the \textit{normal genuine
    topologies}. We would like to point out that dropping the
  requirement that a topology does no become disconnected by the cut
  of a single internal line (1-particle irreducibility), the final
  number of topologies would still be 44.}
\end{table}

\begin{figure}[tbph]
\begin{centering}
\includegraphics[scale=0.90]{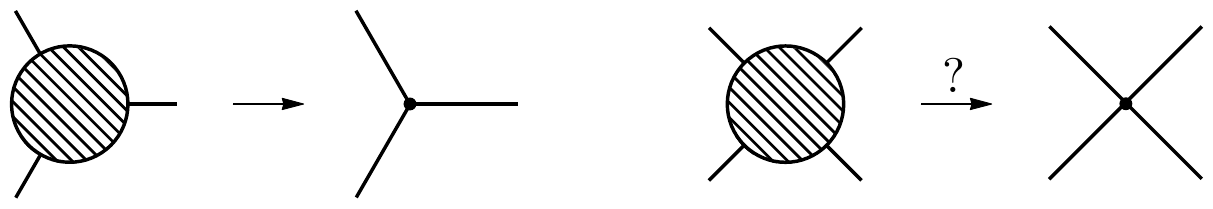}
\par\end{centering}
\protect\caption{\label{fig:loopcontraction}Subparts of diagrams with
  loops and 3 external lines can be compressed into a 3-line vertex,
  reducing the number of loops of the diagram. For subparts with 4
  external lines, this will only generate a renormalizable interaction
  if all external lines are scalars.}
\end{figure}

Consider first 3-point vertices.  No matter what is the particle
content of a model, if a loop with 3 external legs is allowed by
symmetry, so is the trilinear vertex without the loop (see figure
\ref{fig:loopcontraction}). Since this reasoning applies equally to
fermion-fermion-scalar and to scalar-scalar-scalar vertices, this
criterion can be defined at the level of topologies.  For loops with 4
external legs, on the other hand, it is only possible to compress it
to a renormalizable vertex if all external lines are scalars. Thus,
this criterion needs to be used on diagrams, not topologies.  The
important point is that if a diagram has compressible subdiagrams
(with 3 or 4 external legs), it cannot be genuine. We note that there
is also the expectation of the converse: diagrams with incompressible
loops are genuine, as there will be a choice of quantum numbers for
the internal scalars and fermions such that there will be no other
neutrino mass contribution with less loops (in appendix
\ref{sec:appendixB} we present an argument why we believe this is always
true). That is why earlier we called them \textit{normal genuine}
diagrams.

\begin{figure}[tbph]
\begin{centering}
\includegraphics[scale=0.90]{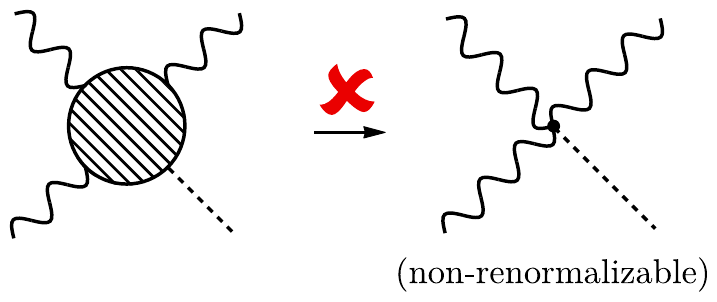}
\par\end{centering} \protect\caption{\label{fig:exception1}Our
  procedure to obtain genuine 3-loop neutrino mass diagrams and
  topologies may not be valid in the presence of vector fields. In
  this example, a 1-loop fragment of a larger diagram cannot be
  contracted into a renormalizable point interaction, see text.}
\end{figure}

The usage of the word "normal" in this context is explained by the
existence of exceptions to the above arguments.  First of all,
strictly speaking, this cut on 4-point vertices is only valid for
diagrams without vector fields.  Consider the diagram shown in figure
\ref{fig:exception1}: It has 3 external vector fields ($V$) and one
scalar ($S$). Yet a term $VVVS$ is not Lorentz invariant, hence such a
loop cannot be compressed into a renormalizable interaction (the
effective interaction is $\partial VVVS$, of dimension 5). As a
consequence of this, some otherwise non-genuine topologies might be
classified as genuine if vector fields are used.  We are mentioning
this exception only for completeness, since we are interested
in diagrams with fermions and scalars only.

\begin{figure}[tbph]
\begin{centering}
\includegraphics[scale=0.90]{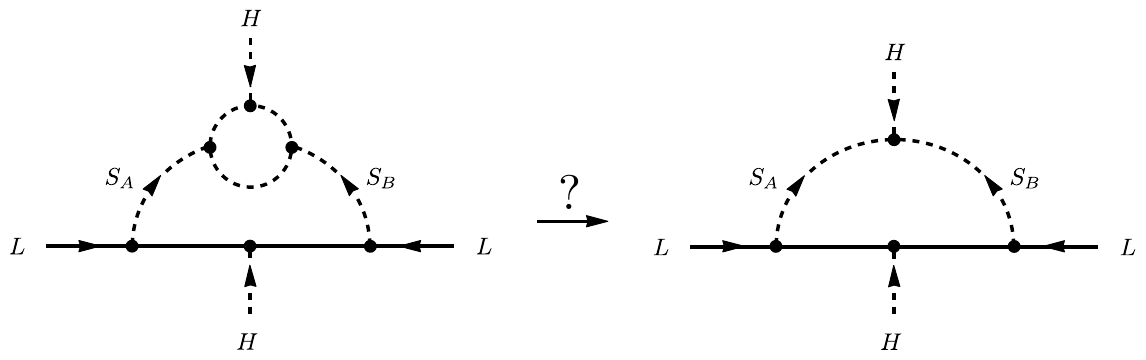}
\par\end{centering} \protect\caption{\label{fig:exception2}This 2-loop
  realization of the dimension five Weinberg operator illustrates a
  loophole in our automatized algorithm for finding genuine $n$-loops
  neutrino mass diagrams and topologies (we therefore track these
  special cases manually). In particular, if the scalar $S_A$ (or
  $S_B$) is the Higgs fields $H$, and $S_B$ (or $S_A$) is an $SU(2)$
  singlet with the correct hypercharge, then there is no point
  interaction $H S_A S_B$. Hence, the existence of the left diagram
  does not imply that one can build the diagram on the right, with one
  loop less.}
\end{figure}

However, a second exception to the procedure used to obtain the
previously mentioned 44 genuine diagrams is more subtle. To understand
it, consider the 2-loop diagram in figure \ref{fig:exception2}. The
diagram appears to be non-genuine because it requires fields with
quantum numbers such that they would have a renormalizable interaction
$H S_A S_B$, which could be used to remove one loop from the
diagram. In a sense, this is indeed always true: such trilinear
combination of fields must be gauge invariant. Yet, $H S_A S_B$ might
be identically zero for specific choices of $S_A$ and $S_B$.  Take
the case where $S_A$ is the Higgs field $H$, and $S_B$ is a scalar
singlet with hypercharge -1. Then, $H H S_B\equiv 0$ because the
$SU(2)$ singlet contraction of two doublets is antisymmetric.  For particular
choices of the quantum numbers of the remaining fields, one can in
fact check that no $d=5$ 1-loop model is generated, hence the 2-loop
diagram/topology in figure \ref{fig:exception2} is in fact
genuine. This construction involving the use of repeated fields to
forbid point-interactions (which otherwise would be allowed by their
quantum numbers) can obviously be extended to 3-loop diagrams. These
genuine 3-loop diagrams (topologies) which lead to non-genuine
model-diagrams, unless very special choices of quantum numbers are
made, we call the \textit{special genuine} diagrams (topologies). Out
of the 160 topologies mentioned above, 44 are normal genuine ones, and
of the remaining 116 there are 29 which fall into this
class.
The complete list is shown in appendix \ref{sec:genuinetopologies},
where we also classify them according to which particular particle
combination is needed to make the corresponding model genuine.

Note that, if we break down the fields into their components, the
neutrino mass obtained from these special topologies arises from a
{\em difference of diagram amplitudes}, with the negative sign(s)
coming from the anti-symmetry of $SU(2)_L$ (and/or color)
contractions. This is very clear, for example, in the 1-loop
subdiagram in figure 4 (on the left), which must correspond to an
$HHS_B$ interaction, as mentioned earlier. In the limit where the
momenta flowing into these critical subdiagrams is small, the
difference of amplitudes will approach zero. However, the momenta
flowing into these subdiagrams is a loop momenta, hence the overall
neutrino mass obtained from special genuine diagrams does not need to
be small when compared to the mass obtained from normal genuine
diagrams.

\begin{figure}[tbph]
\begin{centering}
\includegraphics[scale=0.90]{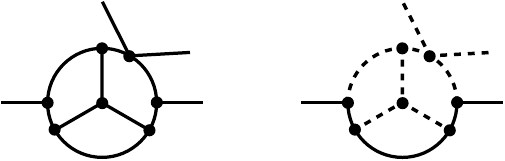}
\par\end{centering} \protect\caption{\label{fig:ngfin}One example of
  a non-genuine but finite topology (to the left), generating the
  diagram on the right. See text.}
\end{figure}

The other 87 remaining topologies ($160=44+29+87$) generate non-genuine
diagrams. Even so, it is important to note that some of these
topologies (27 of them) may lead to \textit{non-genuine finite
  diagrams}. These are diagrams for which an additional (broken)
symmetry is always needed to forbid the otherwise allowed $\ell$-loop
diagrams ($\ell<3$) that result from compressing one or more loops to
a renormalizable vertex. We show an example of such a topology and
corresponding non-genuine finite diagram in figure \ref{fig:ngfin}.
In this diagram, the inner loop on the fermion line is an example of a
compressible 3-point vertex. However, if this fermion is of Majorana
type, one can add to the corresponding model an extra symmetry, for
example a global $U(1)$ as in \cite{Hatanaka:2014tba}.  A SM singlet
scalar can then be coupled to the Majorana fermion and assigned a
charge under the $U(1)$. The tree-level coupling of the compressible
inner loop could be forbidden in this way. Spontaneous breaking of
this $U(1)\to Z_2$ by the vacuum expectation value (vev) of the
singlet generates a Majorana mass term for the fermion and allows then
this 3-loop diagram to exist.

We stress again that we do not consider this class to be genuine, as
these models require extra symmetries (which need to be broken). Note
that the symmetries can not be exact, otherwise the compressible loop
is also forbidden by the symmetry. For the remaining 60 topologies in
this non-genuine class all diagrams have infinite loop integrals. Due
to the large number of topologies in this sub-class, we do not show
them in this paper; they can be can found in \cite{extraData}.

\begin{figure}[tbph]
\begin{centering}
\includegraphics[scale=0.90]{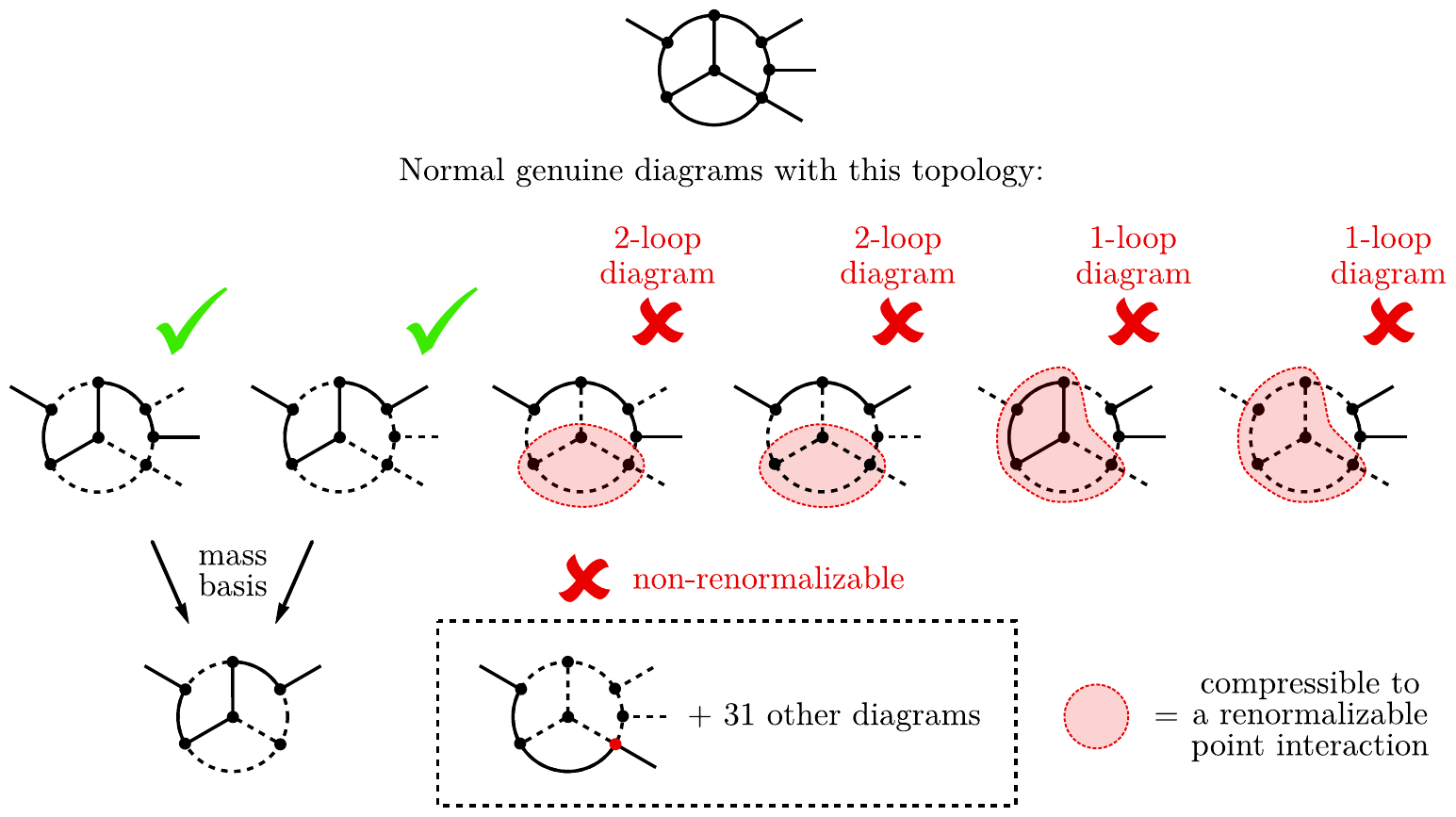}
\par\end{centering}
\protect\caption{\label{fig:example-topology-to-diagrams}There are two
  normal genuine diagrams associated to the topology shown in the top. There
  is also a total of 32 diagrams which can be drawn with a
  non-renormalizable fermion-fermion-scalar-scalar
  interaction. Finally, there are 4 diagrams which are also not
  normal genuine ones because it is possible to shrink a subpart of them into a
  renormalizable point interaction. (Note however that under some very specific circumstances, the third, the fifth and sixth diagrams in the top row can be genuine, hence they are considered \textit{special} genuine diagrams.)}
\end{figure}

We now return to the construction of the genuine diagrams. From the 44
normal genuine topologies, a total of 228 genuine diagrams can be
built \cite{extraData}. In figure
\ref{fig:example-topology-to-diagrams} we show for one particular
topology the possible diagrams, explaining graphically why several of
them are not genuine. Diagrams with compressible loops are discarded
in this set, as well as diagrams with non-renormalizable
interactions. In this particular example, the topology has only two
normal genuine diagrams. The remaining genuine diagrams, of the
special kind, must be found carefully in a non-automated way (there
are 146 of them, 66 of which have a special genuine topology, while
the other 80 have a normal one).

\begin{figure}[tbph]
\begin{centering}
\includegraphics[scale=0.90]{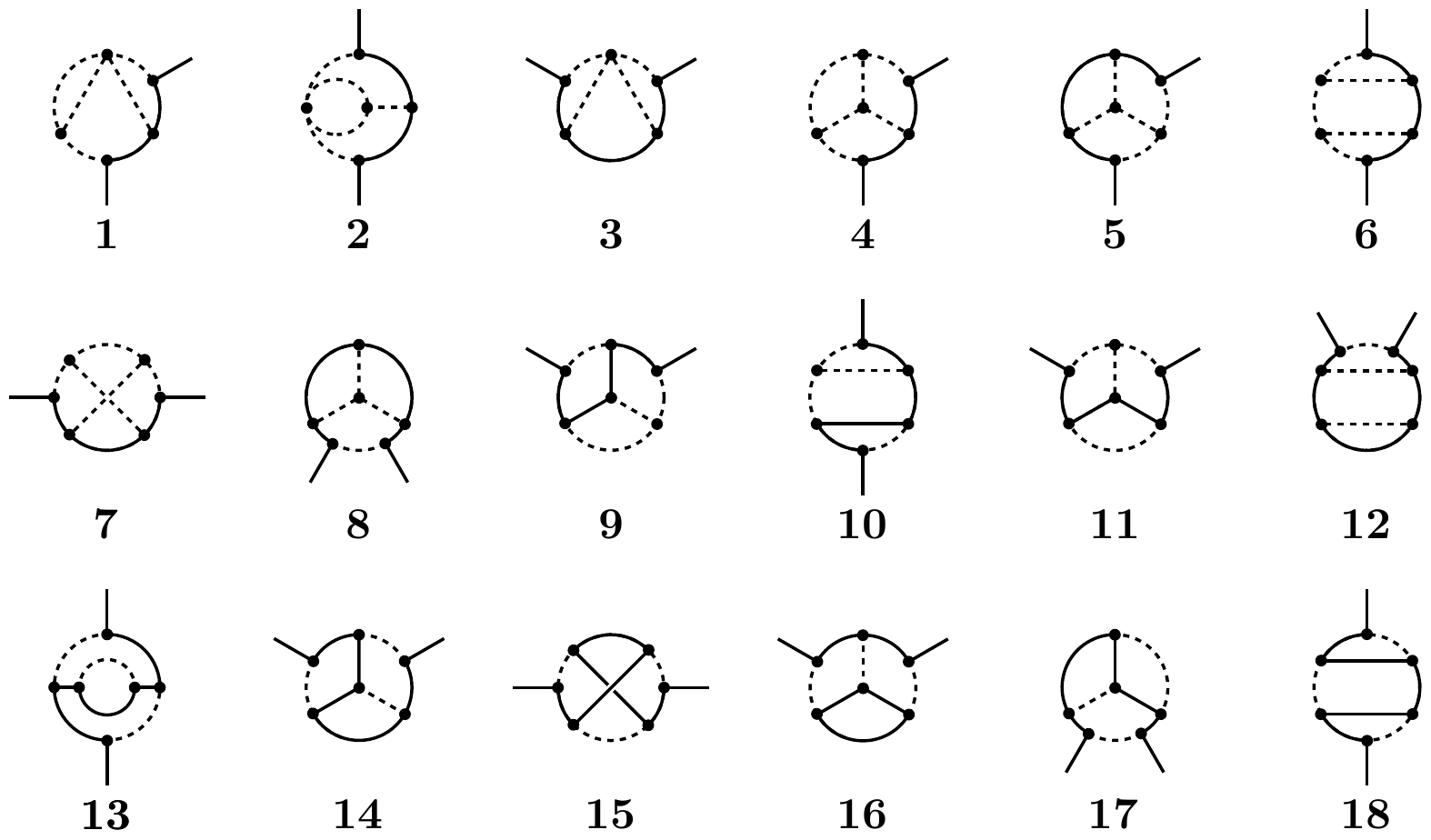}
\par\end{centering} \protect\caption{\label{fig:normalgenuinediagrams}List of
  \textit{normal} genuine diagrams in the mass basis. Note that the two external Higgs lines were
  removed: in general there is a many-to-one relation between the
  original diagrams and the amputated ones shown here. Diagrams in
  this list are referred to in the text as $D^M_i$, where $i$ is the
  number of the diagram shown here. }
\end{figure}

\begin{figure}[tbph]
\begin{centering}
\includegraphics[scale=0.90]{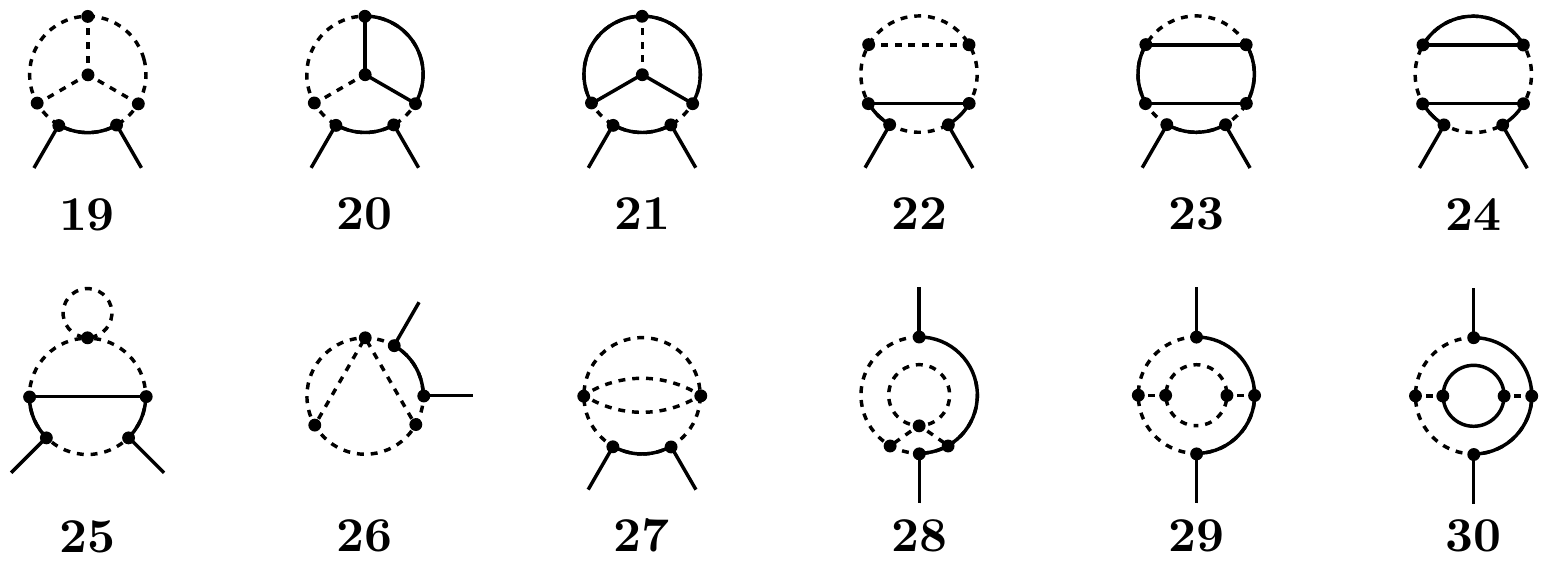}
\par\end{centering}
\protect
\caption{\label{fig:specialgenuinediagrams}List of \textit{special}
  genuine diagrams, with the external Higgs lines removed. Diagrams in
  this list are referred to in the text as $D^M_i$, where $i$ is the
  number of the diagram shown here.}
\end{figure}

\begin{figure}[tbph]
	\begin{centering}
		\includegraphics[scale=0.90]{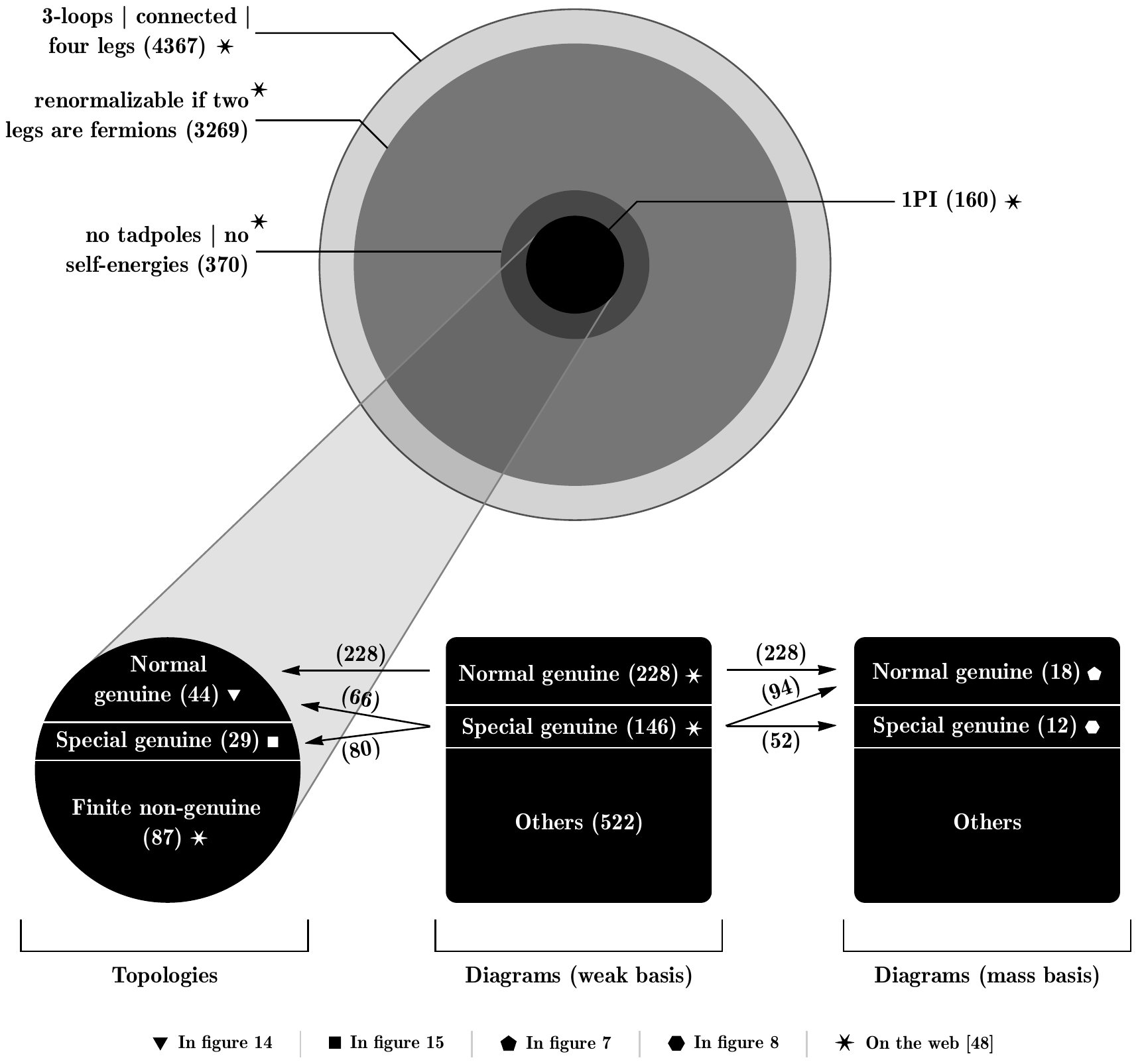}
                \par\end{centering}
                \protect

\caption{\label{fig:summary}Summary of the different types of diagrams
  and topologies. Out of thousands of topologies, only 160 are
  potentially interesting. They correspond to a total of 896 diagrams:
  228 can provide dominant neutrino mass contributions without special
  considerations (\textit{normal genuine diagrams}), and a further 146
  can do so only with very special setups (\textit{special genuine
    diagrams}). We call \textit{normal genuine topologies} to those
  associated to at least one normal genuine diagram (there are 44);
  the \textit{special genuine topologies} are the remaining cases
  which are associated to at least one special genuine diagram (there
  are 29). The remaining topologies are \textit{non-genuine}
  but some of them (27) have at least one finite diagram. Once the external Higgs fields are removed, the
  228 \textit{normal genuine diagrams} become 18 amputated diagrams,
  while the remaining genuine diagrams in the weak basis yield 12 more
  amputated diagrams.}
\end{figure}

As a final step, the two external scalars standing for Higgs vev
insertions are removed, and a list of 18 genuine (amputated) diagrams
is obtained. These are shown in figure \ref{fig:normalgenuinediagrams}. In
other words, the 228 diagrams in the electro-weak basis can be reduced
to 18 diagrams in the mass eigenstate basis. To these one has to add
the 12 diagrams in figure \ref{fig:specialgenuinediagrams} which are
obtained from special genuine diagrams. A visual summary of the steps
described so far, as well as a counting of the genuine diagrams and
topologies, can be found in figure \ref{fig:summary}.  We state again,
that while most of our results apply also to diagrams with vector
bosons, our lists are not complete for vectors, due to the loophole
discussed above in fig.  \ref{fig:exception1}.

We close this section by noting that the amplitudes of the 18+12 diagrams from figures
\ref{fig:normalgenuinediagrams} and \ref{fig:specialgenuinediagrams} can be
decomposed as linear combination of five
master integrals \cite{Martin:2016bgz}. Some of these master integrals admit an
analytical solution, while others can only be solved numerically. A
more detailed discussion about them is given in 
appendix \ref{sec:master integrals}.

\section{Examples}\label{sec:examples}

From the complete set of 228+146 genuine diagrams one can generate models
by assigning quantum numbers to the internal fields following some
basic rules. However, not all will lead to genuine 3-loop neutrino mass models.
For that, one should guarantee the
absence of fields that generate lower order contributions. For example
$\nu_R$, $\Delta$ and $\Sigma$ of the basic three tree-level seesaws,
or the scalar $S\equiv(\mathbf{1},\mathbf{4})_{3/2}$ together with the
fermion $\Psi\equiv(\mathbf{1},\mathbf{3})_{1}$ from the $d=7$
tree-level BNT model \cite{Babu:2009aq}. Here, we introduced the
notation $(\mathbf{SU(3)_C},\mathbf{SU(2)_L})_{Y}$ for the quantum
numbers of internal particles. We will use this notation mostly in the
figures. Note that we shortened this
to $\mathbf{SU(2)_L}_{Y}$ for colourless particles. Thus, for example,
a fermionic $\mathbf{1}_0$ corresponds to a right-handed neutrino
$\nu_R$. As discussed above, it is however possible in many cases to
construct models that avoid lower order diagrams, despite the use of
particles such as $\nu_R$, by adding additional symmetries by hand to
the model. We will show one example of such a model below. In that
case we add a superscript $\omega$ to the particle quantum numbers to
indicate which particles are charged under the new symmetry.  The
simplest possibility is usually a $Z_2$.

\begin{figure}[tb]
\begin{centering}
\includegraphics[scale=1.20]{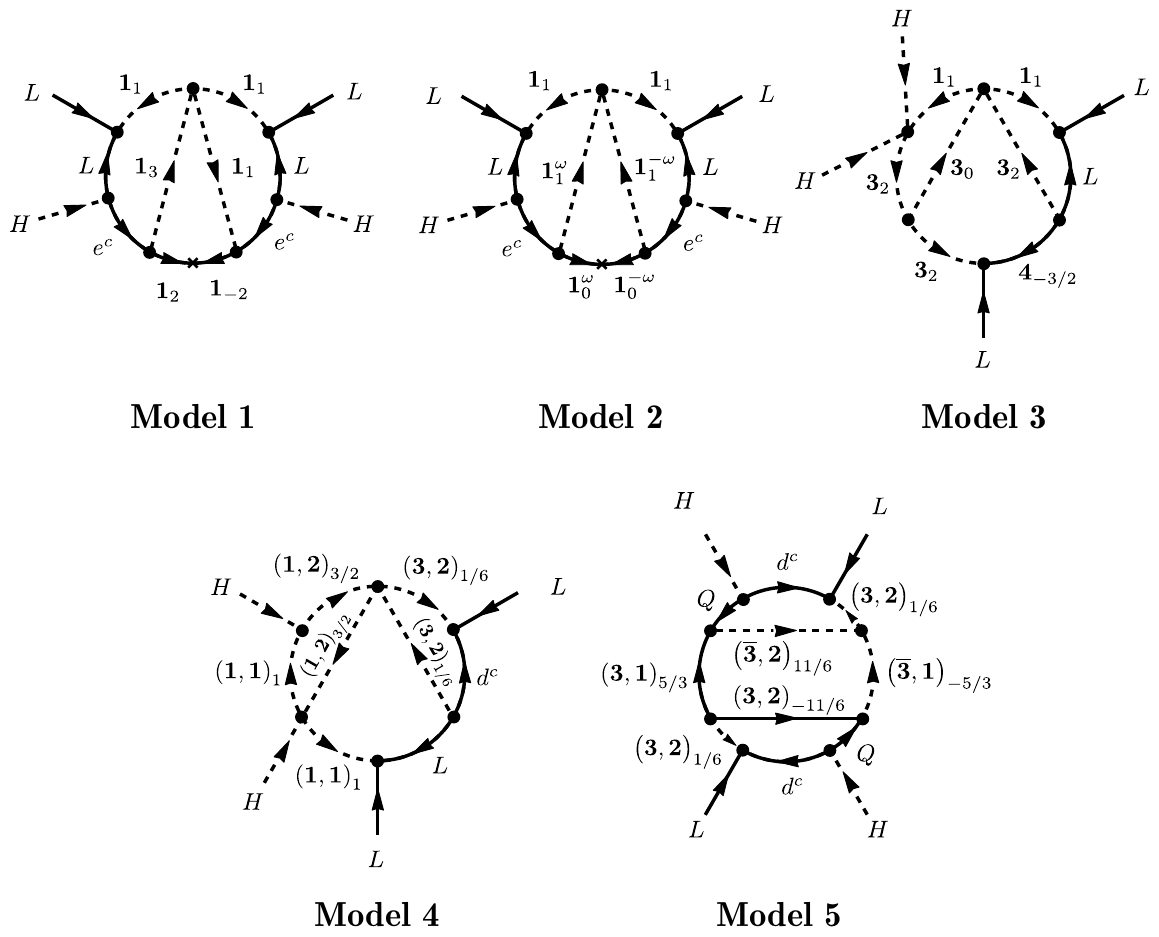}
\par\end{centering}
\protect\caption{\label{fig:model_examples}Five
  examples of three-loop $d=5$ genuine neutrino mass models.}
\end{figure}

Since there are endless possibilities for the quantum numbers of the internal
fermions and scalars, the number of genuine models is infinite.
Here we will just show a few basic examples: Five comparatively
simple models are shown in fig. \ref{fig:model_examples}.  Let us
discuss them briefly.

Model 1, based on the same diagram as the KNT model, can be considered
as one of the simplest genuine 3-loop models possible. The diagram
needs only three singlets (two different scalars and one vector-like
fermion) and no additional symmetry to produce a non-zero neutrino
mass. All other models that we found need either (i) larger $SU(2)_L$
representations and/or (ii) a larger number of beyond SM fields and/or
(iii) an additional symmetry to avoid lower order diagrams.

Model 2 is the simplest realization of the KNT model. It also contains
only three different singlets, as is the case for model 1. However,
the KNT model needs an extra symmetry to avoid a tree-level seesaw
contribution from the fermionic singlet (recall that $\mathbf{1}_0\equiv
\nu_R$). We indicate the particles transforming non-trivially under the
new symmetry, by writing their charge $\omega$ as a superscript.  Note
that for the simplest case of a $Z_2$ this simply reduces to the
particles in the innermost loop to being odd, while the rest of the
diagram contains only particles transforming even. As we mentioned earlier
in the introduction of the paper, there is a number of variations of this diagram in the literature
containing larger $SU(2)_L$ representations in the loops and colored particles as well.

We have chosen model 3 to show how larger $SU(2)_L$
representations can also play a natural role in 3-loop
neutrino mass models. This model is associated to topology $T_{3}$,
being the first 3-loop model to do so in the
literature, as far as we know. There are three new scalars, ${\bf 3}_0$, ${\bf 3}_2$,
${\bf 1}_1$ and one new fermion ${\bf 4}_{-3/2}$ (plus its vector
partner).  The model contains a triply-charged
``leptonic'' fermion as well as a triply charged scalar, and thus it should lead
to a very rich accelerator phenomenology.

The second row in figure \ref{fig:model_examples} shows two
models with coloured fields. Here, we have chosen the simplest
possibilities for colour, i.e. we use only triplets. Variants with
larger colour representations could be created in a straightforward
manner. In the diagram of model 4, colour runs only in one of the loops. This
model has again only three new fields. However, in contrast to models
1--3, here all new fields are scalars. The scalar
$(\mathbf{3,2})_{1/6}$ is a lepto-quark, thus standard LHC
searches for these particles should put bounds on this model. Note that model
4 descends from our topology $T_1$ (we have not found
any model with this topology in the literature). Model 5 is
a second example with coloured particles: it needs
5 exotic fields, but no additional symmetry. Note again that the
exotic fermions in this model, $({\bf 3},{\bf 2})_{-11/6}$ and
$({\bf 3},{\bf 1})_{5/3}$, both must be vector-like.

In the following we will discuss models 1 and 5 in more detail,
including a numerical calculation of the relevant 3-loop integrals. We
will only consider the unrealistic case where one neutrino is massive,
adding just one generation of every new field for
simplicity. Therefore, our results should be understood as estimates
of the typical scale of the neutrino masses and not as a prediction
for their exact values.
Note, however, that it is possible to fit all neutrino oscillation data, including
the mixing angles, in radiative models. Usually, adding more copies of the
exotic fermions is enough (we discuss this briefly at the end of this section).

Also, unless we say otherwise, in the following all dimensionless
couplings are set to one and, in this simplified setup, we
will not put a hierarchy nor flavour structure in the indices of the
Yukawa couplings. (This is done for simplicity; it is not a requirement/constraint on the models.)
When there are no analytical solutions, the calculations for the
three-loop integrals have been done numerically with the code
\texttt{pySecDec} \cite{Borowka:2017idc}. For detailed definitions of the loop integrals see
the appendix \ref{sec:master integrals}.

\subsection{Model 1}\label{subsec:model1}

\begin{table}[tbph]
	\centering
	\begin{tabular}{cccc}
		\hline \hline
		Fields & $SU(3)_C$ & $SU(2)_L$ & $U(1)_Y$  \\
		\hline
		$S_1$ & 1 & 1 & 1
		\\
		$S_2$ & 1 & 1 & 3
		\\
		$F$ & 1 & 1 & 2
		\\
		\hline \hline
	\end{tabular}
\caption{Quantum number assignments for the beyond-the-SM fields of
  model 1 (compare to figure \ref{fig:model_examples}).}
\label{table:model1}
\end{table}

Model 1 contains the SM fields plus the ones given in table
\ref{table:model1}. The fermion $F$ has a vector partner
$\overline{F}$, which is not explicitly shown in the table.  The neutrino mass
in model 1 is generated from the following terms in the Lagrangian:
\begin{align} \label{eq:lagrangian model1}
  \mathscr{L} & = \mathscr{L}_{SM} + Y_1 \overline{L^c} L S_1
        +  Y_2 \overline{F} e^c S_1
        +  Y_3 F e^c S_2^\dagger  +  \lambda_S S_2 (S_1^\dagger)^3  +  \textrm{h.c.}
	\nonumber
	\\
	&+  m_1^2 S_1^\dagger S_1   +  m_2^2 S_2^\dagger S_2
        +  M \overline{F} F + \cdots.
\end{align}
Other quartic terms in the scalar potential (such as $H^\dagger H
S^\dagger S$) are not explicitly given here, as they will only result
in uninteresting corrections to the scalar masses. It is worth
mentioning that $Y_1$ in eq. \eqref{eq:lagrangian model1}, in principle, is a
$3 \times 3$ antisymmetric matrix. This fact is important if one wants
to fit the complete neutrino oscillation data (see the discussion at the end of this section).

The mass diagram of model 1 in figure \ref{fig:model_examples} shows
that the neutrino mass is proportional to the product of two masses of
SM charged leptons. Considering the dominant contribution with two
$\tau$ leptons running in the loop, the neutrino mass matrix is then
calculated straightforwardly as:
\begin{equation} \label{eq:NuMassModel1}
  (M_\nu)_{\alpha \beta} = -\frac{3!}{(16\pi^2)^3} \lambda_S
        \frac{ m_\tau^2 }{M} 
	\left[ (Y_1)_{\alpha \tau} (Y_2)_\tau (Y_3)_\tau (Y_1)_{\tau \beta} \;
          +  \; (\alpha \leftrightarrow  \beta) \right]
	\; F_{loop} \! \left( x_1, x_2 \right).
\end{equation}

After EWSB, in the mass eigenbasis, model 1 generates the diagram $D_3^M$
in figure \ref{fig:normalgenuinediagrams} with a mass insertion in each of
the three internal fermions. From the diagram, and assigning
momenta to the internal fields, we get the following expression for the loop function $F_{loop}$, which is given by a
dimensionless integral:
\begin{eqnarray} \label{eq:Floop model1}
	F_{loop} \! \left( x_1, x_2 \right) = \!\!\!\!\!\! \iiint\limits_{(k_1,k_2,k_3)} \!\!\! \frac{ 1 }{ [k_1^2] [k_1^2-x_1] [k_2^2] [k_2^2-x_1] [k_3^2-1] [(k_2-k_3)^2-x_1] [(k_3-k_1)^2-x_2] }.
\end{eqnarray}
Here $m_\tau$ was neglected, while the other
masses were normalized to the vector-like mass $M$ of the field $F$:
\begin{equation}
	x_1 = \frac{m_1^2}{M^2},\; x_2 = \frac{m_2^2}{M^2}.
\end{equation}
We also used the short-hand notation
\begin{equation} \label{eq:int notation}
	\int\limits_k \equiv (16 \pi^2) \int \frac{ d^4 k }{(2\pi)^4}.
\end{equation}
For the full decomposition of $F_{loop} \! \left( x_1, x_2 \right)$,
in terms of master integrals, suitable for numerical evaluation, 
see appendix \ref{sec:master integrals}.

\begin{figure}[tb]
\begin{centering}
  \includegraphics[scale=0.6]{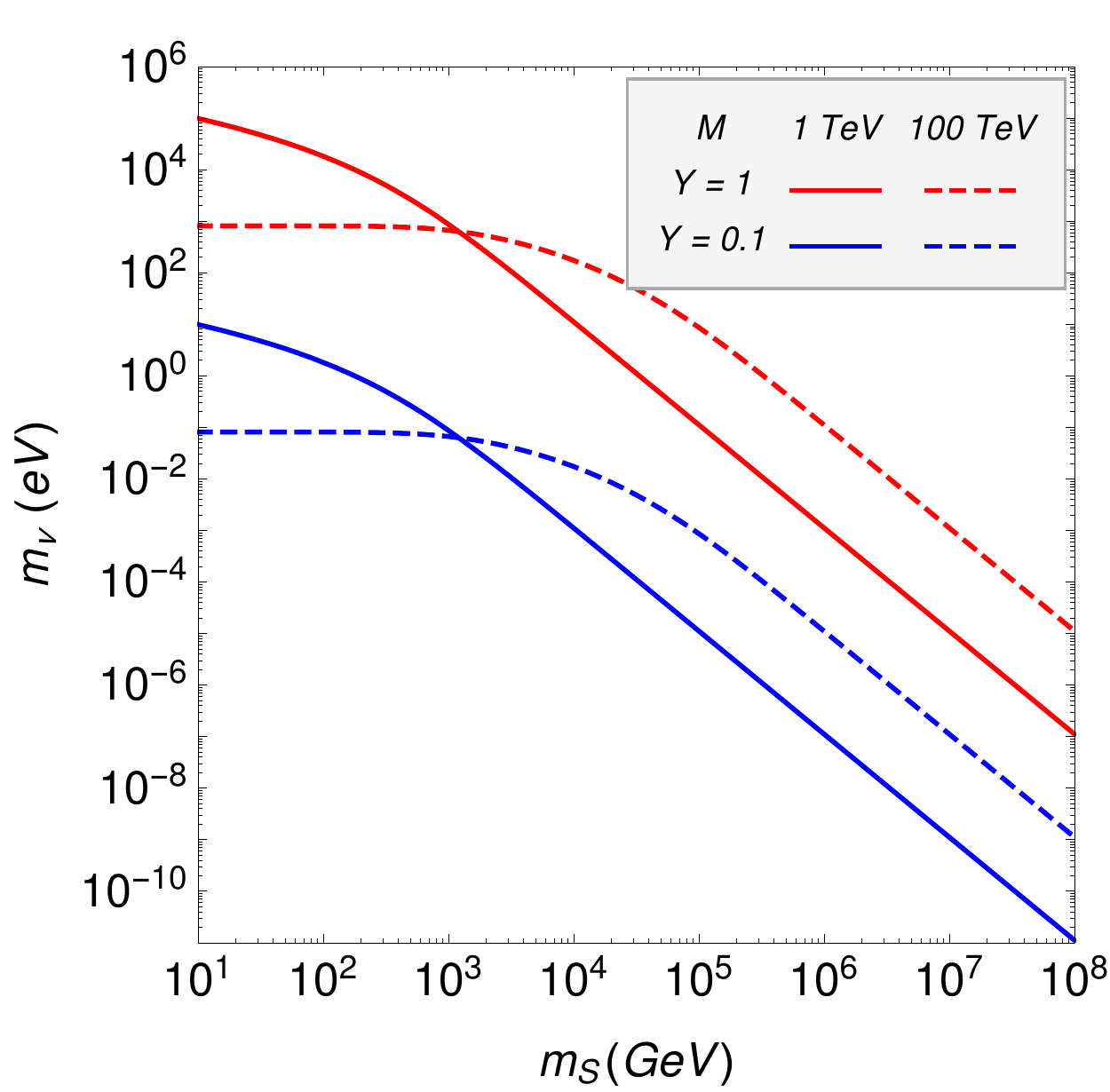}
  \par\end{centering}
\protect\caption{\label{fig:NumassModel1} The neutrino mass scale in
  model 1 for a few sample choices of parameters, see text for
  details.}
\end{figure}

In figure \ref{fig:NumassModel1} we show the neutrino
mass scale for different choices of parameters. For the calculation we
have taken all masses of the new scalar singlets equal, i.e. $m_1 =
m_2 = m_S$. As can be seen from eq.  (\ref{eq:NuMassModel1}), $M_\nu$
is proportional to the fourth power of the Yukawas. For
masses of order $\mathcal{O}(1)$ TeV one can reproduce the neutrino atmospheric scale ($\sim 0.05$ eV) with Yukawas
$\mathcal{O}(10^{-2}-10^{-1})$.

The dependence of the neutrino mass on the masses of the fields in the
loop is also understood straightforwardly. From the diagram of model
1 in the mass eigenbasis, it is straightforward to see that
the neutrino mass should scale as:
\begin{equation} \label{eq:Mnu approx model1}
    M_\nu \sim m_{\tau}^2 \frac{M}{\Lambda^2}, 
\end{equation}
where $\Lambda$ is some characteristic energy scale. As the loop
function contains only two mass scales, i.e. $m_S$ and $M$ (neglecting
$m_\tau$), for $m_S \gg M$ the neutrino mass decreases with $1/m_S^2$,
while for small scalar masses one obtains a constant value (for a fixed
$M$).\footnote{This is true only if
  $M\gg m_{\tau}$. It can be easily checked that for
  $M \rightarrow 0$ the integral vanishes and neutrinos remain
  massless. }

In summary, as figure \ref{fig:NumassModel1} shows, the correct
neutrino mass scale is obtained in this model
for a wide range of masses. In one extreme case, the new physics scale can be as high as
$10^3$ TeV if all Yukawas are
order one. On the other hand, even for masses
$M$ and $m_S$ of the order of $1$ TeV, Yukawa couplings
can be as large as ${\cal O}(0.1)$.

\subsection{Model 5}\label{subsec:model5}

We have performed an analogous study for model 5 of figure
\ref{fig:model_examples}. In the mass eigenbasis, the neutrino diagram
corresponds to diagram 10 in figure \ref{fig:normalgenuinediagrams} with a
mass insertion on both d-quark internal lines.

\begin{table}[tbph]
	\centering
	\begin{tabular}{ccccc}
		\hline \hline
		Fields & $SU(3)_C$ & $SU(2)_L$ & $U(1)_Y$  \\
		\hline
		$S_Q$ & 3 & 2 & 1/6
		\\
		$S_1$ & 3 & 1 & 5/3
		\\
		$S_2$ & 3 & 2 & -11/6
		\\
		$F_1$ & 3 & 1 & 5/3
		\\
		$F_2$ & 3 & 2 & -11/6
		\\
		\hline \hline
	\end{tabular}
\caption{Quantum numbers of the new fields given in model 5 (see
  figure \ref{fig:model_examples}).}
\label{table:model5}
\end{table}

The new fields present in the model are listed in table \ref{table:model5}.
Among others, the Lagrangian contains the following interactions:
\begin{align} \label{eq:lagrangian model5}
	\mathscr{L} &= \mathscr{L}_{SM} + Y_1 L d^c S_Q + Y_2 Q F_1 S_2 + Y_3 Q F_2 S_1 + Y^L_4 \overline{F_1}\, \overline{F_2} S_Q^\dagger + Y^R_4 F_1 F_2 S_Q + \mu_S S_Q^\dagger S_1^\dagger S_2^\dagger +   \textrm{h.c.}
	\nonumber
	\\
	&+  M_{F_1} \overline{F_1} F_1 + M_{F_2} \overline{F_2} F_2 + m_{S_Q}^2 S_Q^\dagger S_Q + m_{S_1}^2 S_1^\dagger S_1 + m_{S_2}^2 S_2^\dagger S_2  + \cdots.
\end{align}
Additional quartic terms in the scalar potential coupling the new
scalars and the higgs field are not written down explicitly.

Similarly to model 1, the neutrino mass in model 5 is proportional to the
product of two d-quark masses. Thus, one expects the dominant
contribution will be proportional to the mass of the bottom quark
squared:
\begin{align} \label{eq:NumassModel5}
  (M_\nu)_{\alpha \beta} &= - \frac{12 \mu_S}{(16\pi^2)^3}
  \frac{m_b^2}{ m_{S_Q}^2 }
  \left[ (Y_1)_{\alpha b} (Y_2)_b (Y_3)_b (Y_1)_{b \beta} \; +  \; (\alpha \leftrightarrow  \beta) \right]  \\ \nonumber
	& \times \left[ Y^L_4 F_L( x_1,x_2,x_3,x_4 ) + Y^{R*}_4 F_R( x_1,x_2,x_3,x_4 ) \right]
\end{align}
where%
\begin{align} \label{eq:Floop model5}
	\hspace*{-5.mm} F_L( x_1,x_2,x_3,x_4 ) = & \;\;\; \!\!\!\!\!\! \iiint\limits_{(k_1,k_2,k_3)} \!\!\! \frac{ \sqrt{x_1 x_3} }{ [k_1^2] [k_1^2-1] [k_2^2] [k_2^2-1] [k_3^2-x_1] [k_3^2-x_2] [(k_2-k_3)^2-x_3] [(k_3-k_1)^2-x_4] }, 
		\\
	\hspace*{-5.mm} F_R( x_1,x_2,x_3,x_4 ) = & \;\;\; \!\!\!\!\!\! \iiint\limits_{(k_1,k_2,k_3)} \!\!\! \frac{ \slashed{k_3}(\slashed{k_2}-\slashed{k_3}) }{ [k_1^2] [k_1^2-1] [k_2^2] [k_2^2-1] [k_3^2-x_1] [k_3^2-x_2] [(k_2-k_3)^2-x_3] [(k_3-k_1)^2-x_4] },
\end{align}
are two dimensionless loop integrals normalized to the mass of the new
scalar $S_Q$,
\begin{equation}
	x_1 = \frac{M_{F_1}^2}{m_{S_Q}^2},\; x_2 = \frac{m_{S_1}^2}{m_{S_Q}^2},\; x_3 = \frac{M_{F_2}^2}{m_{S_Q}^2},\; x_4 = \frac{m_{S_2}^2}{m_{S_Q}^2}.
\end{equation}
For the decomposition of both integrals in terms of master integrals
we refer again to appendix \ref{sec:master integrals}.

\begin{figure}[tb]
\begin{centering}
\includegraphics[scale=0.6]{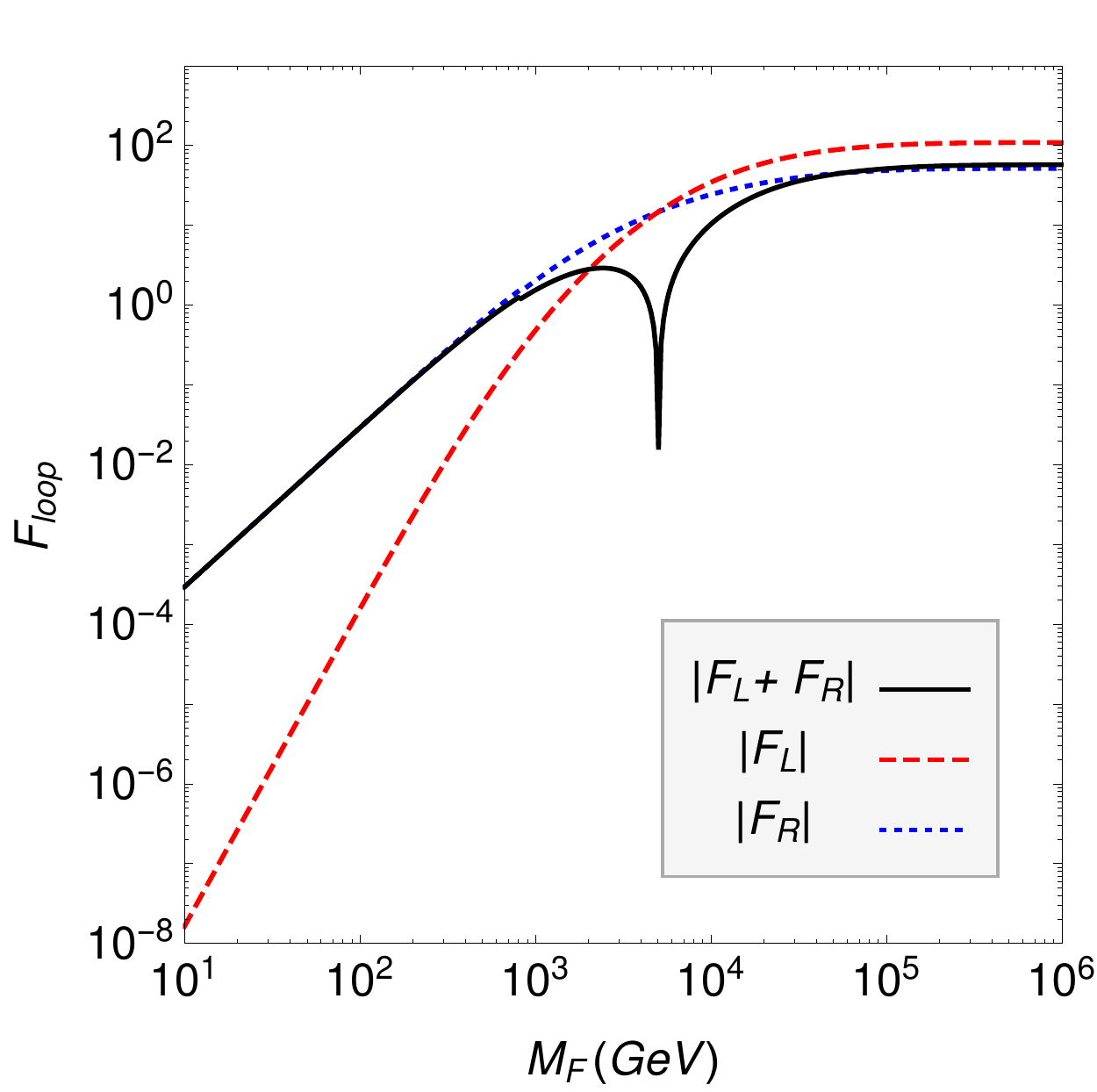}
\par\end{centering} \protect\caption{\label{fig:Loops model5} Loop
  functions, see eq. \eqref{eq:Floop model5}, that enter the neutrino
  mass, see eq. \eqref{eq:NumassModel5}, generated by model 5, see figure
  \ref{fig:model_examples}. For this plot, the masses of $S_1$ and
  $S_2$ are taken to be $1$ TeV, while both fermion masses $M_{F_1} =
  M_{F_2} = M_F$.}
\end{figure}

There are two different integrals contributing to the neutrino mass,
as can be seen in eq. \eqref{eq:NumassModel5}, due to the fact that 
one may flip the chirality of the internal fermions with mass insertions.
$F_1$ and $F_2$ must have vector-like masses\footnote{New fermions beyond
	the standard model fields must have vector-like mass terms for
	phenomenological reasons. For example, a fourth chiral family is
	excluded by the Higgs production measurements at the LHC.}, thus
there are two possible choices for the chiral structure of the vertex,
i.e. either one uses $Y_4^L \overline{F_1}\, \overline{F_2}$
or $Y_4^{R*} (F_1)^*(F_2)^*$. This yields one loop integral with $M_{F_1}
M_{F_2}$ in the numerator and another one with the loop momenta of
both fermions instead of their masses. This fact is important because,
unlike in model 1, for model 5 a cancellation can occur between both
contributions, as shown in figure \ref{fig:Loops model5}.  This
cancellation occurs when all the masses in the diagram are of
the same order. For example, $\mathcal{O}(1)$ TeV in the case shown in
figure \ref{fig:Loops model5}.

\begin{figure}[tb]
\begin{centering}
  \includegraphics[scale=0.6]{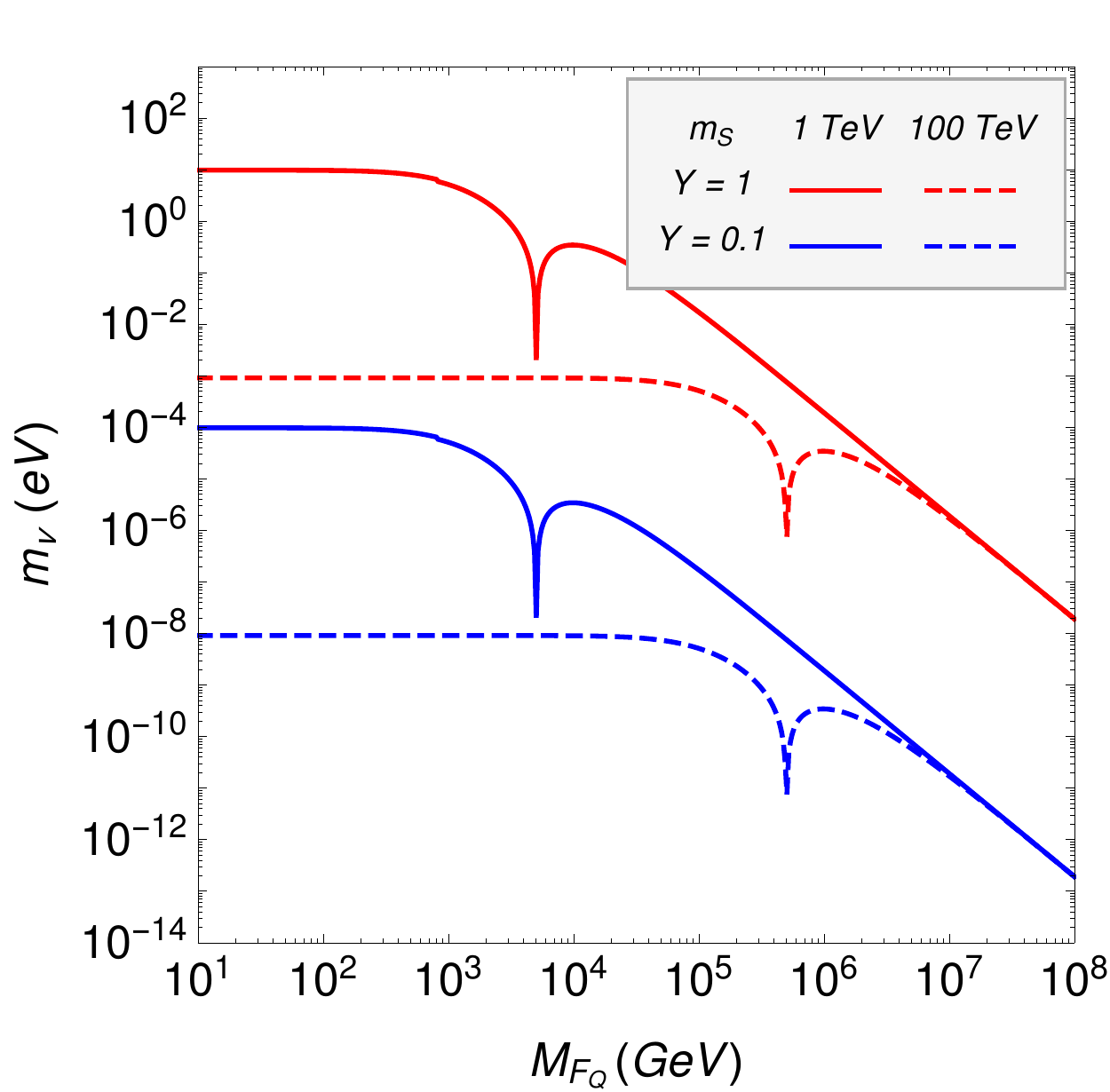}
  \par\end{centering}
\protect\caption{\label{fig:NumassModel5} The neutrino mass scale for
  model 5 in figure \ref{fig:model_examples}, for some example choices
  of parameters. See text for details.}
\end{figure}

In figure \ref{fig:NumassModel5} we show some examples for the
neutrino mass scale for specific but arbitrary choices of
parameters. Taking all the masses of the new scalars equal for
simplicity, i.e. $m_{S_1} = m_{S_2} = m_S$, for masses of
$\mathcal{O}(1)$ TeV Yukawas around ${\cal O}(0.5)$ are needed to
generate the atmospheric scale. The difference, compared to the
previous case of model 1, arises from the fact that the loop integral
in model 5 contains one extra propagator compared to model 1, as well
as one extra Yukawa coupling. Thus, the neutrino mass scales differently in
model 5. The dependence on the masses is again easily understood,
considering that for model 5 one has:
\begin{equation} \label{eq:Mnu approx model5}
    M_\nu \sim \mu_S m_b^2 \left( \frac{M_{F}^2}{\Lambda^4} + \frac{1}{\Lambda^2} \right).
\end{equation}
For $M_{F} \ll m_S$ one has a plateau whose height scales as
$1/\Lambda^2$, instead of $1/\Lambda$ as in model 1, see
eq. \eqref{eq:Mnu approx model5}, while for large fermion masses both models
have the same behaviour.

It is worth mentioning that model 5 contains an extra mass scale, i.e.
the coupling $\mu_S$ in eq. \eqref{eq:lagrangian model5}. In fig.
\ref{fig:NumassModel5} we have chose $\mu_S = m_S$.  Increasing its
value will smoothen the differences between both models, making it possible 
to reach the measured neutrino mass scale with smaller Yukawas couplings. 
On the other hand, the need of at least one
neutrino mass of the order of $0.05$ eV can be interpreted
as a lower limit on this parameter.

We have chosen to discuss models 1 and 5 in more detail because
they span the typical range of three-loop neutrino mass
models. By direct inspection of the genuine diagrams, listed in figure
\ref{fig:normalgenuinediagrams}, it can be seen that every integral contains
7 or 8 propagators, leading to the same behaviour as in either model 1
or model 5, respectively, in the limit of large masses. For small
scalar and fermion masses, the scale of $M_\nu$ depends on the
numerator of the integral, i.e. the number of fermions inside the loop
along with the chiral structure of each vertex, as well as the
presence of SM mass insertions.

Finally, we should point out that obviously any realistic neutrino
mass model should be able to reproduce all neutrino oscillation data,
i.e. the two neutrino mass squared differences along with three
neutrino mixing angles and phases. The aim of our simplified examples
was to show how the neutrino mass scales in typical 3-loop models; it
was not to make a thorough neutrino flavour fit. However, going beyond
the simplified scenario where there is just one non-zero charged
lepton (or down-quark) mass the neutrino mass matrices given in
\eqref{eq:NuMassModel1} and \eqref{eq:NumassModel5} have rank-2. This
makes it possible to fit normal or inverted hierarchical neutrino
spectra, including a correct fit for angles and phases, in both model
1 and model 5.  In order to fit a degenerate neutrino spectrum, a
rank-3 neutrino mass matrix is needed. This can be achieved easily in
model 5 just by adding extra copies of the new fields, for instance
having two copies of $F_1$ and $F_2$. However, fitting a degenerate
spectrum is not possible for the case of model 1, disregarding the
number of copies of the fields.  This is due to the antisymmetry of
the Yukawa $Y_1$.\footnote{This can be understood recalling that the
  rank of any $n \times n$ antisymmetric matrix is at most $n-1$ for
  odd $n$'s, together with the identity $\text{rank}(AB) \leq
  \text{min}( \text{rank}(A), \text{rank}(B) )$, for two arbitrary
  matrices $A$ and $B$.} Again, as with the overall mass scale, our
two example models represent the two typical kind of models, that can
be found at 3-loop order.

\section{Conclusions}

In this paper we have discussed the complete decomposition of the
Weinberg operator at 3-loop order. Our analysis concentrates on
finding those topologies and diagrams
that can give the dominant contribution to the neutrino mass
matrix, without the use of additional symmetries beyond those
of the standard model. We call such topologies/diagrams genuine.
We considered models with scalars and fermions only.

The requirement of ``genuineness'' eliminates the large majority of
possible topologies: From more than four thousands, there are only 73 topologies
which satisfy this criteria. We have discussed how to
identify these cases and we listed them in
appendix \ref{sec:genuinetopologies}. Those 73 genuine
topologies  were sub-divided into two classes: Normal ones (44 topologies)
and special ones (29 topologies). While the former can be
found systematically by our selection criteria, the latter topologies form
an exception to our general rules,
as explained in detail in section \ref{sect:class} and in the appendix
\ref{sec:genuinetopologies}. 
This exception is related to the fact that usually,
if any three fields (or four scalars) can interact through a loop, then
they can also do so through a renormalizable local interaction.
However, for special combinations of fields this is is not true:
for example, the Higgs-Higgs-singlet local interaction is null, but a 
loop with these 3 external scalars does not need to have a zero amplitude.

The 44 topologies we have found are associated to a total of 228 diagrams in the
electro-weak basis, from which one can get 3-loop leading order neutrino
masses contributions. Going to the mass eigenstate basis, this list is reduced
to only 18 diagrams (they are shown in figure \ref{fig:normalgenuinediagrams}). To these normal genuine diagrams one has to add 146 special ones
, in the electroweak basis, which give another 12 mass eigenstate
diagrams (see figure \ref{fig:specialgenuinediagrams}) We have also
discussed how all those diagrams can be calculated with only
five master integrals which where analysed in the literature previously \cite{Martin:2016bgz}. We give them in
appendix \ref{sec:master integrals}, where we also show how the loop integrals for specific neutrino mass
models can be constructed with two examples.

We have then also shown in section \ref{sec:examples}, how our general
results can be easily used to build genuine 3-loop neutrino mass
models. A few examples are briefly mentioned, and for two of them we have calculated
 the neutrino mass scale in more detail. This allows us to estimate the typical
parameter range (couplings and masses), for which these 3-loop models
can explain the measured neutrino oscillation data. We find that dimension 5 3-loop models
will give a good fit to data if the new particles have masses
roughly in the range $1-10^3$ TeV.  Such a low scale is partially
testable at current and future colliders, as well as in experiments searching
for lepton flavour violation.  Thus, 3-loop models are interesting
constructions, since they are experimentally testable. We hope that
model builders will find our results useful.

\bigskip

\centerline{\bf Acknowledgements}

\medskip
This work was supported by the Spanish grants FPA2017-85216-P and
SEV-2014-0398 (AEI/FEDER, UE), Spanish consolider project MultiDark
FPA2017-90566-REDC, FPU15/03158 (MECD) and PROMETEOII/2018/165 
(Generalitat Valenciana).  R.F. also acknowledges the financial
support from the Grant Agency of the Czech Republic, (GA\v{C}R),
contract nr. 17-04902S, as well as from the grant Juan de la
Cierva-formaci\'on FJCI-2014-21651 (from Spain).

\bigskip
  
\appendix

\section{\label{sec:genuinetopologies}List of genuine topologies}

\begin{figure}[tbph]
\begin{centering}
\includegraphics[scale=0.85]{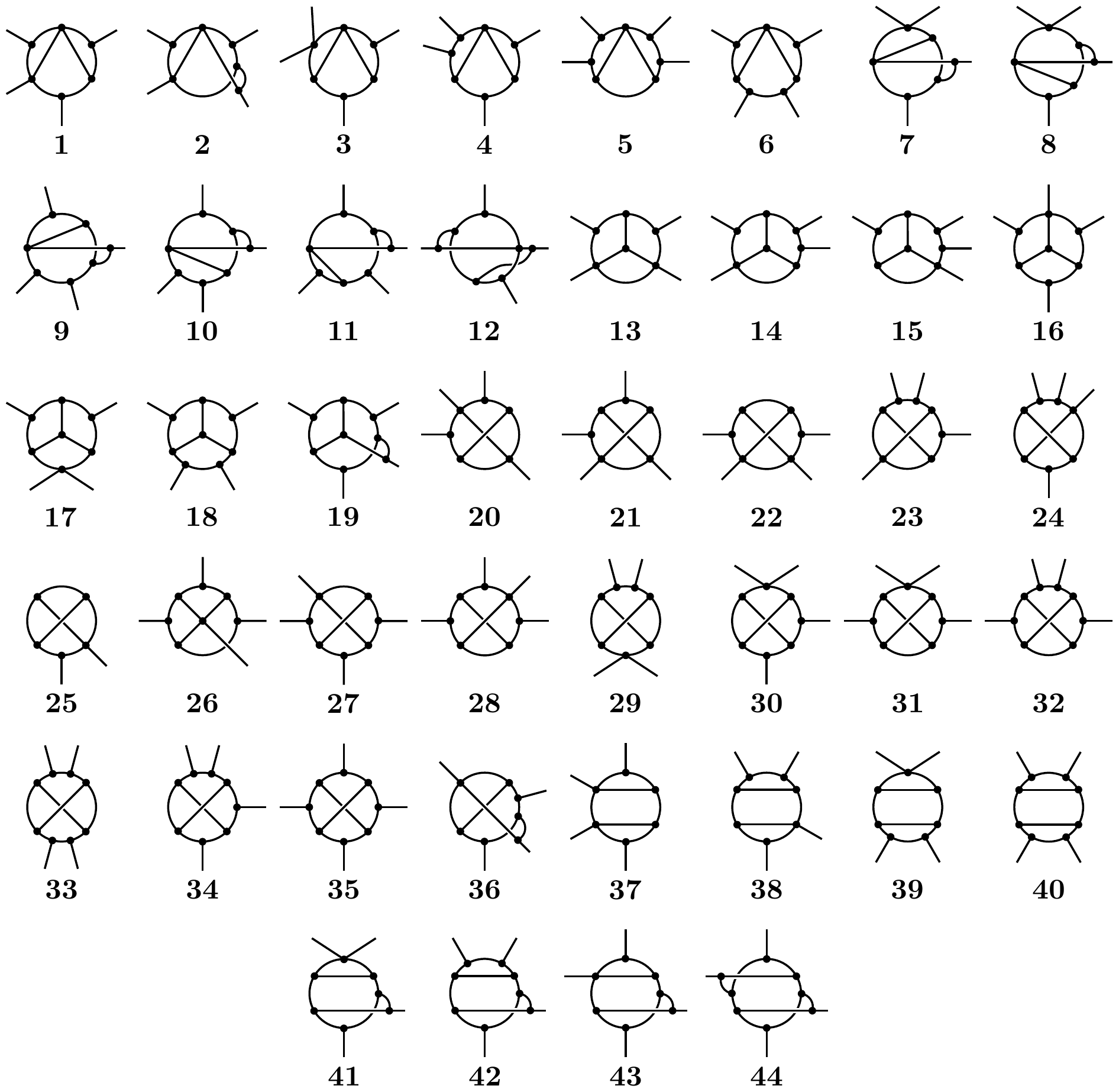}
\par\end{centering}

\protect\caption{\label{fig:topologies_normal}List of topologies
  associated to \textit{normal} genuine diagrams. We refer to them in the text as $T_i$ with $i=1,\cdots,44$.}
\end{figure}

In this work we are interested in those scenarios where the dominant
contribution to neutrino masses arises from a 3-loop realization of
the Weinberg operator. As explained in the main text, genuine neutrino mass
diagrams must descend from one of the 44 topologies shown in figure
\ref{fig:topologies_normal}, otherwise it is not possible to forbid
lower order contributions, independently of the assignment of fermions
and scalars to the lines.

\begin{figure}[tbph]
	\begin{centering}
		\includegraphics[scale=0.85]{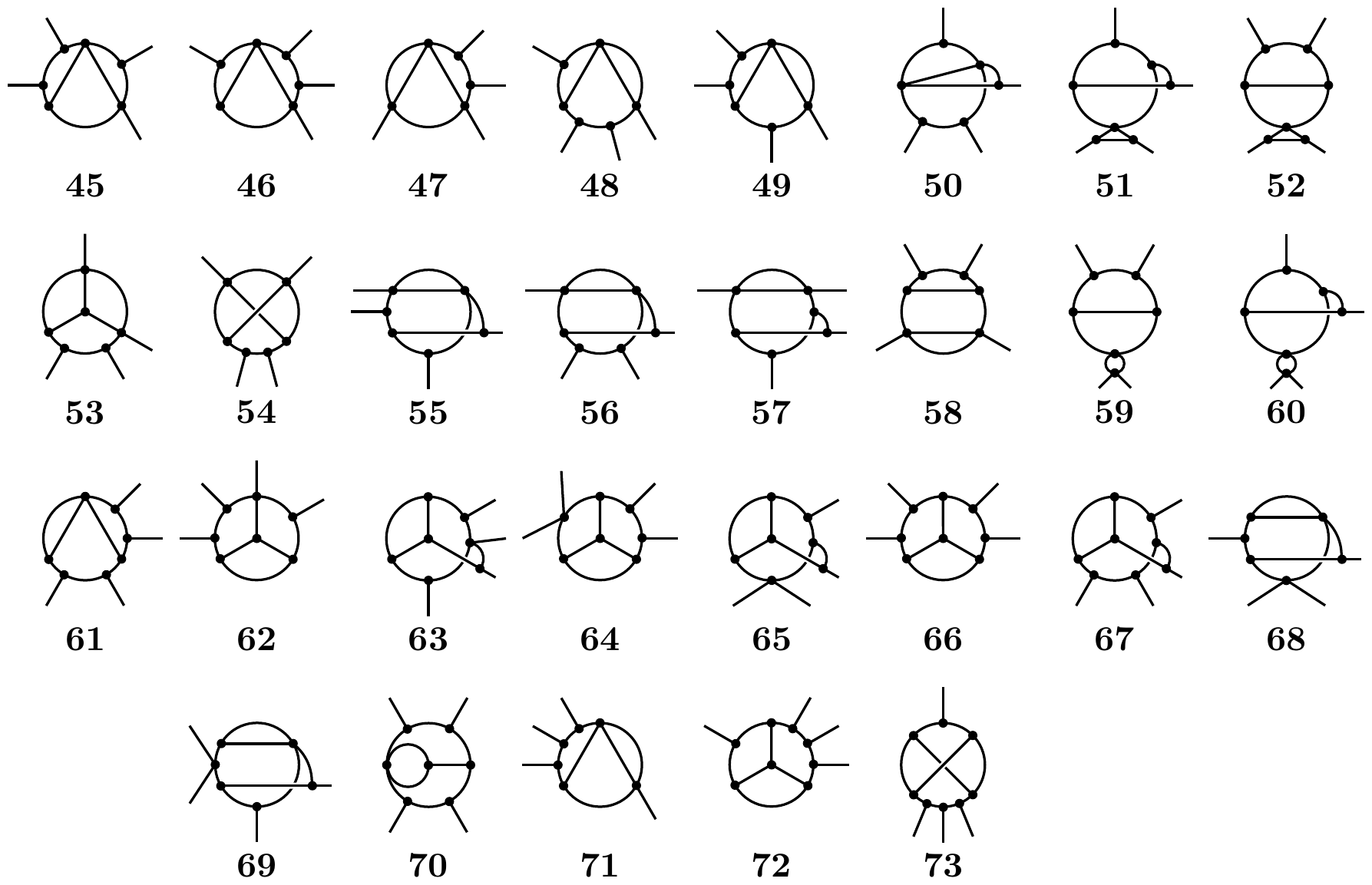}
		\par\end{centering}
	
	\protect\caption{\label{fig:topologies_special}List of topologies associated to \textit{special} genuine diagrams ($T_i$ with $i=45,\cdots,73$).	 The topologies are ordered according to specific field content needed in order to be considered as genuine. See the text for details.}
\end{figure}

However, the procedure used to identify these 44 \textit{normal}
genuine topologies admits a loophole: in the presence of very special
fields, it is possible to generate 3-loop neutrino masses diagrams
with other topologies, with no lower order contributions appearing. In
figure \ref{fig:topologies_special} we show these 29 \textit{special}
genuine topologies.

\begin{figure}[tbph]
	\begin{centering}
		\includegraphics[scale=1.5]{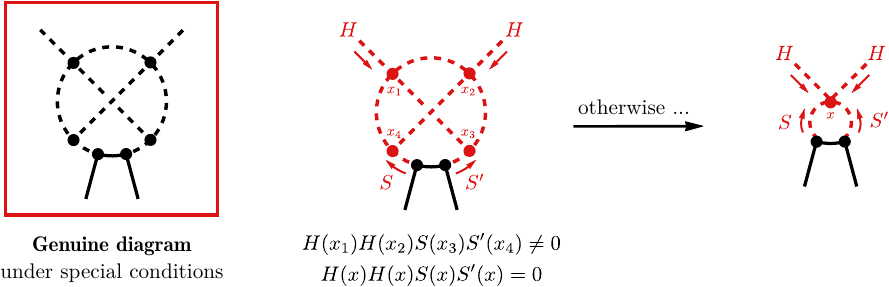}
		\par\end{centering}
	
	        \protect
\caption{\label{fig:topologies_special_example1}Topology 54 has only
  one diagram associated to it (shown here inside the box on the
  left). This diagram is only genuine under special conditions, in
  particular the two scalars interacting with the fermion line ($S$
  and $S'$) must be a Higgs $H$ and a scalar $\phi_D\equiv
  \left(\boldsymbol{1},\boldsymbol{2},-3/2\right)$.  }
\end{figure}

Consider topology 54: there is only one way of making a fermion chain
connecting the two external $L$'s hence there is a single diagram to
be considered (see figure \ref{fig:topologies_special_example1}). One
can identify in it a 2-loop subdiagram with 4 external scalar lines,
shown in red in the middle of figure
\ref{fig:topologies_special_example1}. Two of the external scalars are
the Higgs fields of the Weinberg operator, while the others ($S$ and
$S^\prime$) are unknown a priori, hence the subdiagram is associated
to the operator $HHSS'$. This means that, for most assignments of
quantum numbers to the internal fields, one can write down such an
interaction directly in a renormalizable Lagrangian, in which case
neutrino masses can be generated via the 1-loop diagram shown in
figure \ref{fig:topologies_special_example1} on the right.

However, strictly speaking the 2-loop subdiagram generates the
non-local operator $H(x_1)H(x_2)S(x_3)S'(x_4)$ which we may rewrite as
\begin{equation}
H\left(x_{1}\right)H\left(x_{2}\right)S\left(x_{3}\right)S'\left(x_{4}\right)=H\left(x\right)H\left(x\right)S\left(x\right)S'\left(x\right)+\sum_{n}c_{n}\frac{\partial^{2n}}{\Lambda^{2n}}H\left(x\right)H\left(x\right)S\left(x\right)S'\left(x\right)\,,
\end{equation}
where $x$ is some space-time point close to the $x_i$,
$\Lambda$ is some mass scale and the $c_n$ are adimensional
parameters. If $H(x)H(x)S(x)S'(x)$ is nullified, the 1-loop neutrino
mass depicted in figure \ref{fig:topologies_special} will not
exist. Keeping in mind that in the Weinberg operator the two Higgs
doublets contract as an $SU(2)$ triplet, there is only one possibility
of avoiding a renormalizable $HHSS'$ interaction: setting $S=H$ and
$S'=\phi_D\equiv \left(\boldsymbol{1},\boldsymbol{2},-3/2\right)$, or
vice-versa.\footnote{If $S^\prime$ was a $SU(2)$ quadruplet,
$S'= \left(\boldsymbol{1},\boldsymbol{4},-3/2\right)$, the local
operator $H(x)H(x)H(x)S'(x)$ would not vanish. Another idea to avoid
the $HHSS'$ point interaction is to have
$S=S'=\left(\boldsymbol{1},\boldsymbol{R},-1/2\right)$ for some
$SU(2)$ representation $\boldsymbol{R}$, such that the two $S$'s
contract antisymmetrically. This happens for odd-dimensional $SU(2)$
representations (other than the trivial one):
$\boldsymbol{R}=\boldsymbol{3},\boldsymbol{5}, \boldsymbol{7}, \cdots
$. The problem is that, for hypercharge $-1/2$, such
$\boldsymbol{R}$'s lead to fractionally charged particles, the
lightest of which would be stable and therefore pose a cosmological
problem \cite{Langacker:2011db} (adding a non-trivial colour quantum
number would not change this). Therefore we discarded such scenarios
altogether.} With this very special setup, the 3-loop
diagram in figure \ref{fig:topologies_special_example1} is genuine,
and that is why the corresponding topology is included in figure
\ref{fig:topologies_special}.

\begin{figure}[tbph]
	\begin{centering}
		\includegraphics[scale=1.5]{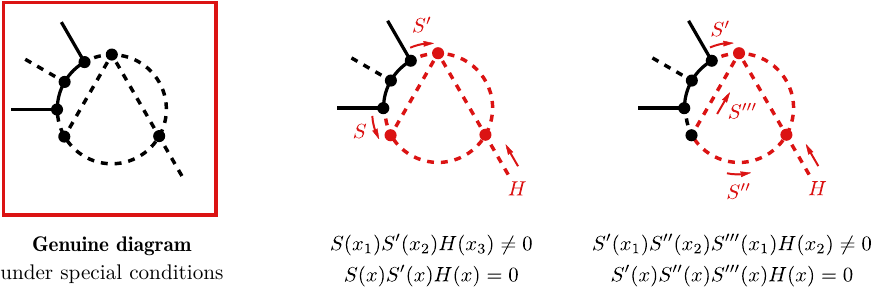}
		\par\end{centering}
	
	\protect\caption{\label{fig:topologies_special_example2}Topology 71 has one genuine diagram, shown here. The quantum numbers of the scalars must be special, otherwise one could build the Weinberg operator from a 1- or 2-loop diagram. In particular, this scenario is avoided only if $S=\phi_S\equiv \left(\boldsymbol{1},\boldsymbol{1},-1\right)$, $S^\prime=H$, $S^{\prime\prime}=\phi_{y}\equiv\left(\boldsymbol{1},\boldsymbol{1},y\right)$, and $S^{\prime\prime\prime}=\widetilde{\phi}_{y}\equiv\left(\boldsymbol{1},\boldsymbol{1},-1-y\right)$ for some hypercharge $y\neq 0, \pm 1, \pm 2$.}
\end{figure}

As a more involved example, we will now discuss topology 71, for which
there is a single genuine diagram, shown in figure
\ref{fig:topologies_special_example2}. In this same figure, we
indicate in red two subdiagrams (with 3 and 4 external lines) which
should not be shrinkable to point interactions, otherwise the diagram
becomes non-genuine. In particular, the internal scalars must be such
that the local operators $HSS^\prime$ and $HS^\prime S^{\prime\prime}
S^{\prime\prime\prime}$ are zero, while still allowing their non-local
counterparts.

To nullify the first interaction, $HSS^\prime$, we must have either $S^\prime=S$, $S=H$ or $S'=H$. But the first possibility ($HSS$) is no good, as there is no $SU(2)$ representation $\boldsymbol{R}$ such that $\boldsymbol{R}\times \boldsymbol{R} \times \boldsymbol{2}$ is gauge invariant. The second and third possibilities ($HHS$ and $HHS'$), on the other hand, imply that either $S'=\phi_S\equiv \left(\boldsymbol{1},\boldsymbol{1},-1\right)$ or $S=\phi_S$, respectively.

We consider now the other interaction, $HS^\prime S^{\prime\prime} S^{\prime\prime\prime}$, which also needs to be zero in its point-like realization. Given the two possible quantum number assignments for $S'$, we might have $H\phi_S S'' S'''$ or $HH S'' S'''$. However, it is not complicated to check that $H\phi_S S'' S'''$ would require either $S''$ or $S'''$ to be a gauge singlet $\left(\boldsymbol{1},\boldsymbol{1},0\right)$, so one could make a 2-loop realization of the Weinberg operator by removing this scalar line from the 3-loop diagram.

We then proceed with the only viable hypothesis --- $HS^\prime S^{\prime\prime} S^{\prime\prime\prime}=HH S'' S'''$. Again, we are faced with two scenarios: (a) one of the undetermined scalars ($S^{\prime\prime}$ and $S^{\prime\prime\prime}$) is equal to $H$, or (b) both $S^{\prime\prime}$ and $S^{\prime\prime\prime}$ are different from $H$. Scenario (a) implies that $\left(S^{\prime\prime},S^{\prime\prime\prime}\right)=\left(H,\phi_D\right)$, while scenario (b) leads to $S^{\prime\prime}=\phi_{y}\equiv\left(\boldsymbol{1},\boldsymbol{1},y\right)$, and $S^{\prime\prime\prime}=\widetilde{\phi}_{y}\equiv\left(\boldsymbol{1},\boldsymbol{1},-1-y\right)$ for some $y$. In the last case, both scalars must be $SU(2)$ singlets in order to ensure that the field product $S^{\prime\prime}S^{\prime\prime\prime}$ does not have a triplet component which would be responsible for coupling the two $H$'s symmetrically.

\begin{figure}[tbph]
	\begin{centering}
		\includegraphics[scale=1.5]{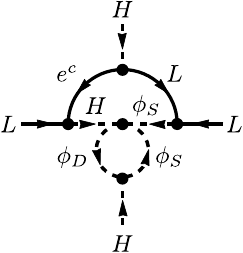}
		\par\end{centering}
	
	\protect\caption{\label{fig:phiD-phiS-diagram}Two loop diagram which can be built in a model with the scalar fields $\phi_S\equiv \left(\boldsymbol{1},\boldsymbol{1},-1\right)$ and $\phi_D\equiv \left(\boldsymbol{1},\boldsymbol{2},-3/2\right)$. Note that because $H(x)H(x)\phi_S(x)=0$, it is not possible to remove the bottom loop in the diagram. }
\end{figure}

Taking into consideration everything said so far, it might then seem
that there are two possibilities for topology 71 with the labelling as
indicated on figure \ref{fig:topologies_special_example2}:
$\left(S,S',S'',S'''\right)=\left(\phi_{S},H,H,\phi_{D}\right)$ and
$\left(\phi_{S},H,\phi_{y},\widetilde{\phi}_{y}\right)$ (possibly
switching the quantum number of $S''$ and $S'''$). However, a model
with both the scalar $\phi_S$ and the scalar $\phi_D$ will inevitable
generate the 2-loop diagram shown in figure
\ref{fig:phiD-phiS-diagram}, so the diagram in figure
\ref{fig:topologies_special_example2} is genuine only if
$\left(S,S',S'',S'''\right)=\left(\phi_{S},H,\phi_{y},\widetilde{\phi}_{y}\right)$. It
is worth mentioning that although the scalar loop with $\phi_{S}$ and
$\phi_{D}$ in figure \ref{fig:phiD-phiS-diagram} seems to diverge, the
loop is finite. This is because such special diagrams involve
differences of two diagrams due to the $SU(2)_L$ contractions,
removing the divergences. This is the same contractions that makes precisely
$H(x)H(x)\phi_S(x)=0$.

For all the topologies in figure (\ref{fig:topologies_special}), we
performed a similar analysis as in the previous examples, identifying
the loop or loops at the diagram level that can exploit the loophole
and, therefore, the specific field content needed for the diagram to
be genuine. With this analysis, we classified the topologies in three
groups such that all the diagrams generated by a certain topology
require the same fields to be genuine. In figure
(\ref{fig:topologies_special}), the first two rows of topologies (from
topology 45 to 60) generate diagrams which contain one or two 4-point
loop scalar vertices with at least one external higgs. The models
descending from these topologies necessarily have the fields $\phi_y$,
$\widetilde{\phi}_y$ and/or $\phi_D$ in order to be
genuine. Topologies 61 to 70 generate diagrams with one 3-point
internal loop, i.e. with no leg being a external leptons or
Higgs. This can be either a 3-point scalar or fermion-fermion-scalar
vertex. Note that in both cases the tree level should be zero, so one
cannot have more than one copy of these fermions or scalars. Finally,
the diagrams coming from the last three topologies 71, 72 and 73,
contain at least one reducible loop with two scalars and an external
higgs. In principle, one can avoid the corresponding 2-loop diagram
with the recipe described in figure
(\ref{fig:topologies_special_example2}), thus making the topology
genuine. Nevertheless, we mention that this diagram alone generates
models which are not able fit neutrino data as they contain a 
structure identical to the simplest realization of the Zee model
\cite{Zee:1980ai}.\footnote{Note that unlike the Zee model where
  another copy of the Higgs can be added to fit neutrino data, in the
  case under discussion this is not a viable solution, because the
  resulting model will be a correction to a dominant 2-loop model
  generated by reducing the 3-point scalar loop with the copy of the
  Higgs.}

Just like the two example above, many of the 29 special topologies in figure \ref{fig:topologies_special} admit only 1 valid diagram, with specific quantum numbers assigned to some of the internal lines. Indeed, there is a total of 80 diagrams associated to these 29 topologies, which are distributed as follows (for topologies 45 to 73): 1, 1, 1, 1, 2, 3, 1, 1, 1, 1, 1, 4, 2, 2, 4, 2, 20, 18, 1, 1, 1, 1, 2, 3, 1, 1, 1, 1, 1. The complete list of diagrams can be found in \cite{extraData}.

\section{\label{sec:appendixB}Relation between incompressible loops and genuineness of a diagram}

We have mentioned in section \ref{sect:class} that those diagrams for
which it is possible to compress one or more loops into a
renormalizable vertex $v$ are not genuine\footnote{We have also
  discussed in detail an exception to this rule, due to the potential
  presence of repeated fields. Hence we introduced the concept of
  special genuine diagrams and topologies, which are genuine even
  though they have compressible loops.}, as one can then use the
interaction $v$ to construct a similar diagram with less loops. In
other words,

\begin{equation}
\textrm{loop compressibility }\Rightarrow\textrm{non-genuineness}\,.
\end{equation}
Obviously, this is equivalent to the statement that genuine diagrams
have incompressible loops ($\textrm{genuineness}\Rightarrow\textrm{loop incompressibility}$).
However, this is not the same as 
\begin{equation}
\textrm{loop incompressibility }\Rightarrow\textrm{genuineness}\label{eq:implication}
\end{equation}
and yet it was stated before that we expect this to be true. Indeed,
our analysis relies on this important assumption, so in this appendix
we discuss why we believe it to be true. We think that the argument
presented here is compelling, but we stop short of calling it a proof.

First, consider the following intuitive/informal explanation for the
implication (\ref{eq:implication}).  For particular assignments of
quantum numbers to the internal lines of a diagram, there might be
extra interactions between some of the diagram's fields which are
completely unrelated to the interactions used in the diagram. If that
is the case, it might be possible to construct the Weinberg operator
$LLHH$ with less loops by using the additional interactions.  However,
there will be a choice of quantum numbers of the internal lines such
that this does not happen: no extra ``non-trivial interactions'' (see
below) between the fields is possible, hence the operator $LLHH$
cannot be realized by a simpler diagram, with less loops.

\begin{figure}[tbph]
\begin{centering}
\includegraphics[scale=0.7]{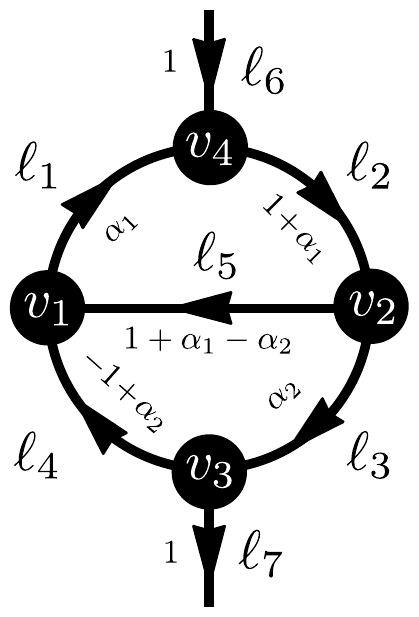}
\par\end{centering}

\protect\caption{\label{fig:Two-loop-diagram}Two loop diagram with oriented lines
$\ell_{i}$ and vertices $v_{j}$. The hypercharge of each line is
indicated as a function of two free numbers: $\alpha_{1}$ and $\alpha_{2}$.
The hypercharge of the external lines was fixed to 1 in this example. }

\end{figure}

In order to formalize this idea, consider only the abelian $U(1)_{Y}$
symmetry. In a $n$-loop diagram where the hypercharge of the external
particles is fixed, the hypercharges $y_{i}$ of the internal lines
depend on $n$ free numbers $\alpha_{j}$.  More specifically, the
$y_{i}$ are linear functions of these $n$ parameters:
\begin{equation}
y_{i}=c_{0}^{i}+\sum_{j=1}^{n}c_{j}^{i}\alpha_{j}\,,\label{eq:yi}
\end{equation}
where the $c_{0}^{i}$ and $c_{j}^{i}$ are numbers which depend on the
hypercharge of the external particles and on the
topology.\footnote{Given that the $\alpha_{j}$ are free numbers, there
  is some arbitrariness in the choice of $c_{j}^{i}$: for any
  invertible matrix $X$ one can replace $c_{j}^{i}$ by
  $\sum_{j'}X_{jj'}c_{j'}^{i}$.} Figure \ref{fig:Two-loop-diagram}
shows an example where $y_{6}=y_{7}=1$ by choice, and
$y_{1}=\alpha_{1}$, $y_{2}=1+\alpha_{1}$, $y_{3}=\alpha_{2}$,
$y_{4}=-1+\alpha_{2}$, $y_{5}=1+\alpha_{1}-\alpha_{2}$ .

A crucial question is then the following: what is the full list of interactions between the 
fields used in the diagram? From the point of view of the $U(1)_{Y}$
symmetry, any hypothetical interaction beyond those used in the diagram will either be (a) forbidden,
(b) allowed for particular values of the $\alpha_{j}$ or (c) allowed
for all values of the $\alpha_{j}$. Referring to figure (\ref{fig:Two-loop-diagram}),
$\ell_{3}\ell_{4}^{*}$, $\ell_{1}\ell_{2}\ell_{3}$ and $\ell_{1}^{*}\ell_{2}\ell_{7}^{*}$
(respectively) are examples of each of these interactions. We will
only be interested in those interaction of type (c) because we can
choose the $\alpha_{j}$ in order to build a model where all interactions
of type (a) and (b) are forbidden.

For the rest of this discussion, it is important to keep in mind that
the $U(1)_{Y}$ symmetry is blind to combinations of a field and its
conjugate, $\ell_{i}\ell_{i}^{*}$, hence one can add/remove them
at will from any allowed vertex. Now, note that the hypercharges
$y_{i}$ in equation (\ref{eq:yi}) are the most general solutions
to a linear system of equations
\begin{equation}
\sum_{j}C_{ij}y_{j}=0\,,\label{eq:matrixC}
\end{equation}
where the rows of the matrix $C$ represent each vertex, and its columns
stand for each line in the diagram: if line number $j$ enters(leaves)
vertex $i$, then $C_{ij}=1$($-1$), otherwise this entry is null.
For the example in figure \ref{fig:Two-loop-diagram} we would have
the following matrix:
\begin{equation}
C=\left(\begin{array}{ccccccc}
-1 & 0 & 0 & 1 & 1 & 0 & 0\\
0 & 1 & -1 & 0 & -1 & 0 & 0\\
0 & 0 & 1 & -1 & 0 & 0 & -1\\
1 & -1 & 0 & 0 & 0 & 1 & 0
\end{array}\right)\,.
\end{equation}
For each external line in the diagram, since its hypercharge is fixed,
one must add that constraint as well.

The important point is that any new vertex would correspond to adding
a new row to matrix $C$. This operation will not change the solution
space if and only if the new row is a linear combination of the
existing rows (the solutions of $C\cdot y=0$ and $C'\cdot y=0$ are the
same only if the lines of $C'$ are linear combinations of those of
$C$).  Additions and subtractions of rows of $C$ translates into
making new vertices $v$ which are the product of existing ones
(addition) or its conjugates (subtraction):
$v=v_{i}^{(*)}v_{j}^{(*)}v_{k}^{(*)}\cdots$.\footnote{The fact that
  the hypercharge of external lines is fixed introduces a
  complication: the previous statement is true, but one can also add
  hyperchargeless combinations of the external fields. In the case of
  external Higgses $H$ and $L$'s, that would correspond to the
  combination $HL$, but one can invoke additionally Lorentz invariance
  to allow only the addition of pairs of this combination,
  i.e. $HHLL$. However, the sum of all vertices
  $v_{1}v_{2}v_{3}\cdots$ in our diagrams, by construction, yield the
  Weinberg operator (times irrelevant combinations of the internal
  lines of the form $I_{i}I_{i}^{*}$), so the original statement
  stands: the only extra vertices which do not spoil the solution
  space in equation (\ref{eq:yi}) are the trivial ones formed from the
  product of the vertices in the diagram and their conjugates.} The
vertices obtained in this way account for all \textit{unavoidable
  interactions} (considering the $U(1)_Y$ group only).

For example, there are vertices $v_{1}=\ell_{1}^{*}\ell_{4}\ell_{5}$
and $v_{2}=\ell_{2}\ell_{3}^{*}\ell_{5}^{*}$ and in figure \ref{fig:Two-loop-diagram},
so the interactions $v_{2}v_{2}=\ell_{2}\ell_{2}\ell_{3}^{*}\ell_{3}^{*}\ell_{5}^{*}\ell_{5}^{*}$
or $v_{1}^{*}v_{2}=\ell_{1}\ell_{2}\ell_{3}^{*}\ell_{4}^{*}\ell_{5}^{*}\ell_{5}^{*}$
cannot be forbidden by any choice of $\alpha_{1},\alpha_{2}$. Nevertheless,
these are non-renormalizable interactions: to reduce the number of
fields in these new interactions, one can only use the fact that combinations
of the form $\ell_{i}\ell_{i}^{*}$ are irrelevant for the $U(1)_{Y}$
symmetry, hence for example $v_{1}v_{2}=\ell_{1}^{*}\ell_{2}\ell_{3}^{*}\ell_{4}$
(modulo $\ell_{5}\ell_{5}^{*}$). Graphically, it is very easy to
follow what is happening: we remove the line $\ell_{5}$ connecting
the vertices $v_{1}$ and $v_{2}$, condensing them into a quartic
interaction. Applying this process repeatedly for the internal lines
$\ell_{3}$, $\ell_{4}$ and $\ell_{5}$, one generates the new interaction
$\ell_{1}^{*}\ell_{2}\ell_{7}^{*}$ ($=v_{1}v_{2}v_{3}$ modulo $\ell_{i}\ell_{i}^{*}$'s)
which graphically is obtained by collapsing the lower loop of the
diagram into a point. 

Another interesting example are those cases where the same field
appears in more than one line in the diagram. For the present
discussion what is important are those situations where this is
\textit{unavoidable}, rather than just possible. According to the
previous discussion, two lines $\ell$ and $\ell^{\prime}$ must have
the same hypercharge (and therefore potentially represent the same
field) if and only if the bilinear interaction $\ell^{*}\ell^{\prime}$
is \textit{unavoidable}, i.e.  one must be able to merge various of
the diagram vertices into such interaction. Lines $\ell_{6}$ and
$\ell_{7}$ in figure \ref{fig:Two-loop-diagram} constitute an example
of such a scenario (they are external lines, hence their hypercharge
was fixed to 1 in our example, but even if they were internal lines of
a bigger diagram, $U(1)_{Y}$ invariance could not forbid the coupling
$\ell_{6}^{*}\ell_{7}$).  \\

In summary,
\begin{enumerate}
	\item For a $n$-loop diagram, the hypercharge of the internal lines depends
	on $n$ free numbers $\alpha_{i}$.
	\item The only interactions between the various lines which cannot be forbidden
	for any choice of $\alpha_{i}$ (the\textit{ $U(1)_{Y}$ unavoidable
		interactions}) are those for which the sum of hypercharges is identically
	0, i.e. $0+0\alpha_{1}+0\alpha_{2}+\cdots0\alpha_{n}$. (Only a subset
	of these interactions are truly unavoidable as one should take into
	account the full Standard Model symmetry, as well as Lorentz invariance.)
	\item The full list of \textit{$U(1)_{Y}$ unavoidable interactions} can
	be obtained by merging together the diagram vertices (and/or their
	conjugates). In this merging process, $\left(\textrm{line X}\right)\left(\textrm{line X}\right)^{*}$
	combinations can be added or removed. 
	\item Most of these \textit{unavoidable interactions} are non-renormalizable.
	Renormalizable new vertices can be formed only if there are those
	removable $\left(\textrm{line X}\right)\left(\textrm{line X}\right)^{*}$
	combinations mentioned earlier, otherwise by merging vertices
	the number of lines constantly increasing. Graphically, this corresponds
	to coalescing adjacent vertices in the diagram, and removing the line(s)
	connecting them.\footnote{Even though it is not important for the present discussion, we mention
		here that the deleted lines can be the external $L$'s and/or $H$'s: graphically
		one would connect the diagram with a conjugated copy of itself (that
		is, a copy of the diagram with the orientation of all lines flipped)
		via the external $L$ and/or $H$ lines, which would become internal lines.
		For example, consider a diagram with Higgs interactions $v_{1}=H\ell_{1}\ell_{2}$
		and $v_{2}=H\ell_{3}\ell_{4}$. Then $v_{1}v_{2}^{*}=\ell_{1}\ell_{2}\ell_{3}^{*}\ell_{4}^{*}$
		is a new, unavoidable interaction which can be obtained graphically
		in the way just described.} 
\end{enumerate}
There are 4 external lines ($LLHH$) indirectly connected to each
other through a web of vertices and internal lines. The \textit{unavoidable}
alternative ways of connecting these 4 lines must be through an alternative
web of \textit{unavoidable} vertices. Graphically, this new web of
lines and vertices must be obtained from the original one by the vertex-merging
process described above. So, if it is not possible to remove one
or more loops from a diagram by collapsing them into a point, the
diagram is genuine.

\section{Master integrals for 3-loop neutrino masses} \label{sec:master integrals}

In this appendix, we give the minimal set of integrals that span the
complete list of possible genuine models.
In principle, starting from the list of 30 genuine diagrams in the
mass eigenbasis (see fig. \ref{fig:normalgenuinediagrams} and
\ref{fig:specialgenuinediagrams}), one should obtain at least 30
integrals assigning momenta to the fields. This initial set, however,
can be further reduced applying the results previously used in 2-loop
calculations in \cite{Sierra:2014rxa,McDonald:2003zj}, both based on
\cite{vanderBij:1983bw}, and 3-loop integrals in
\cite{Freitas:2016zmy,Martin:2016bgz}.  Here, we are going to
summarize the results in these papers which we need.

Using the notation of \cite{Martin:2016bgz}, we can write:
\begin{dmath} \label{eq:T def}
    \mathbf{T}^{(n_1,n_2,n_3,n_4,n_5,n_6)} (x_1,x_2,x_3,x_4,x_5,x_6) = \iiint\limits_{(k_1,k_2,k_3)} \frac{1}{ [k_1^2-x_1]^{n_1} [k_2^2-x_2]^{n_2} [k_3^2-x_3]^{n_3} [(k_1-k_2)^2-x_4]^{n_4} [(k_2-k_3)^2-x_5]^{n_5} [(k_3-k_1)^2-x_6]^{n_6} }
\end{dmath}
Here we have used the abbreviation given in eq. \eqref{eq:int notation}
and the powers of the propagators $n_i$ can be any integer
number. Note that the integral is invariant under the interchange of
pairs $(n_i,x_i)$ and moreover, satisfies the nine identities obtained
by integration by parts \cite{Chetyrkin:1981qh},
\begin{equation} \label{eq:partial identities}
    0 = \iiint\limits_{(k_1,k_2,k_3)} \frac{\partial}{\partial k_i^\mu} \left[ k_j^\mu \mathbf{X} \right],
\end{equation} 
for $i,j=1,2,3$, with $\mathbf{X}$ equal to any product of propagators
of the form shown in eq. \eqref{eq:T def}. One can find very useful identities
for this kind of integrals, such as:
\begin{equation} 
    \frac 3 2 d + \sum\limits_{j=1}^6 (x_j \mathbf{j}^+-1) n_j = 0.
\end{equation}
Here $d$ is the dimension of the momentum integration in dimensional
regularization and $\mathbf{j}^\pm$ is short-hand notation for the following operator:
\begin{equation} 
    \mathbf{j}^\pm \mathbf{T}^{(...,n_j,...)} = \mathbf{T}^{(...,n_j \pm 1,...)}.
\end{equation}

By repeated application of the
identities \eqref{eq:partial identities}, any of the
3-loop integrals $\mathbf{T}$ can be reduced to a linear combination of
five master integrals \cite{Martin:2016bgz}:
\begin{align} 
        \nonumber
    \mathbf{H}(x_1,x_2,x_3,x_4,x_5,x_6) &= \mathbf{T}^{(1,1,1,1,1,1)} (x_1,x_2,x_3,x_4,x_5,x_6)
    \\ \nonumber
    \mathbf{G}(x_3,x_1,x_6,x_2,x_5)     &= \mathbf{T}^{(1,1,1,0,1,1)} (x_1,x_2,x_3,x_4,x_5,x_6)
    \\ 
    \mathbf{F}(x_1,x_2,x_5,x_6)         &= \mathbf{T}^{(2,1,0,0,1,1)} (x_1,x_2,x_3,x_4,x_5,x_6)
    \\ \nonumber
    \mathbf{A}(x_1)\mathbf{I}(x_2,x_3,x_5) &= \mathbf{T}^{(1,1,1,0,1,0)} (x_1,x_2,x_3,x_4,x_5,x_6)
    \\ \nonumber
    \mathbf{A}(x_1) \mathbf{A}(x_2) \mathbf{A}(x_3) &= \mathbf{T}^{(1,1,1,0,0,0)} (x_1,x_2,x_3,x_4,x_5,x_6).
\end{align}
where $\mathbf{A}$ is the standard one-loop Passarino-Veltman function
\cite{Passarino:1978jh} and $\mathbf{I}$ is a two-loop integral
described in \cite{vanderBij:1983bw}.  It is worth mentioning that
analytical expressions exist for the well-known integrals
$\mathbf{A}$ and $\mathbf{I}$, while for the three-loop ones ($\mathbf{F}$, $\mathbf{G}$, and $\mathbf{H}$) results are known only for
particular cases (see
\cite{Martin:2016bgz} for details).  \\

Particularizing to our case, starting from the 30 diagrams in figure
\ref{fig:normalgenuinediagrams} and \ref{fig:specialgenuinediagrams}, in the
mass insertion approximation, and assigning momenta to the internal lines, one
can find that the integrals have repeated propagators with equal
momenta but different masses.\footnote{The momenta flowing into the diagrams is set to 0, given the smallness of neutrino masses.} One can prove that every 3-loop integral
in figure \ref{fig:normalgenuinediagrams} and
\ref{fig:specialgenuinediagrams} can be written in terms of the integrals
in eq. \eqref{eq:T def}. We note here that
partial fractions can be used to reduce the number of propagators with common momenta \cite{vanderBij:1983bw}:
\begin{align} \label{eq:propagator reduce}
    \mathbf{T}^{(\left\lbrace n_{11},n_{12} \right\rbrace,n_2,n_3,n_4,n_5,n_6)} ( \left\lbrace x_{11},x_{12} \right\rbrace&, x_2,x_3,x_4,x_5,x_6) = 
    \\ \nonumber
    \frac{1}{x_{11}-x_{12}} \Big[ 
    &\mathbf{T}^{( \left\lbrace n_{11},n_{12}-1 \right\rbrace,n_2,n_3,n_4,n_5,n_6)} ( \left\lbrace x_{11},x_{12} \right\rbrace,x_2,x_3,x_4,x_5,x_6) 
    \\ \nonumber
    -&\mathbf{T}^{( \left\lbrace n_{11}-1,n_{12} \right\rbrace,n_2,n_3,n_4,n_5,n_6)} ( \left\lbrace x_{11},x_{12} \right\rbrace,x_2,x_3,x_4,x_5,x_6) \Big],
\end{align}
where $\mathbf{T}^{(\left\lbrace n_{11},n_{12}\right\rbrace,n_2,n_3,n_4,n_5,n_6)} ( \left\lbrace x_{11},x_{12}
\right\rbrace,x_2,x_3,x_4,x_5,x_6)$ is the same as $\mathbf{T}$ without the braces,
but with an extra propagator $[k_1^2-x_{12}]^{n_{12}}$.

On the other hand, some integrals with a non-trivial integrand numerator
can be further simplificatied using the
 \textit{$p^2$-decomposition}, namely
\begin{equation} \label{eq:p2 decomposition}
    \frac{p^2}{(k^2-x_1)(p^2-x_2)} = \frac{1}{(k^2-x_1)} + \frac{x_2}{(k^2-x_1)(p^2-x_2)}.
\end{equation}

To demonstrate how this procedure works in practice, we can take for
instance the loop integral of model 1, given in eq. \eqref{eq:Floop model1} of
section \ref{sec:examples}. Applying the identity
\eqref{eq:propagator reduce} twice to both propagators sharing
$k_1$ and $k_2$ momenta, $F_{loop} \! \left( x_1, x_2 \right)$ can be directly
decomposed in terms of a linear combination of $\mathbf{G}$'s:
\begin{equation} 
    F_{loop} \! \left( x_1, x_2 \right) = \frac{1}{x_1^2}  \bigg\{ \mathbf{G}(1,x_1,x_2,x_1,x_1) - \mathbf{G}(1,x_1,x_2,0,x_1) - \mathbf{G}(1,0,x_2,x_1,x_1) + \mathbf{G}(1,0,x_2,0,x_1)  \bigg\}.
\end{equation}

For model 5, the decomposition of the loop integral $F_L( x_1,x_2 )$
in \eqref{eq:Floop model5} is straightforward given the previous
example. One only has to apply eq. \eqref{eq:propagator reduce} three
times to obtain a linear combination of eight $\mathbf{G}$
integrals. Here we focus on the decomposition of $F_R( x_1,x_2 )$,
just to present an example of a integral with a non-trivial
numerator. One should first notice that under the integral sign
\begin{equation} 
    \slashed{k_3}(\slashed{k_2}+\slashed{k_3}) \longrightarrow k_3 \cdot (k_2+k_3) = \frac 1 2 \left[ (k_2+k_3)^2-k_2^2+k_3^2 \right].
\end{equation}
It is clear that one should apply the \textit{$p^2$-decomposition}
in eq. \eqref{eq:p2 decomposition} along with the partial fractions decomposition \eqref{eq:propagator reduce},
as in the previous case, to get rid of the numerator and the repeated
propagators. The full process of the decomposition is rather lengthy
and cumbersome, so here we give just the final result.
\begin{align} 
    F_L( x_1,x_2,x_3,x_4 ) = \frac{\sqrt{x_1 x_3}}{x_1-x_2} 
    \bigg\{ 
        &\mathbf{G}(x_1,1,x_4,1,x_3) - \mathbf{G}(x_1,1,x_4,0,x_3) - \mathbf{G}(x_1,0,x_4,1,x_3) + \mathbf{G}(x_1,0,x_4,0,x_3) 
        \nonumber\\
        -&\mathbf{G}(x_2,1,x_4,1,x_3) + \mathbf{G}(x_2,1,x_4,0,x_3) + \mathbf{G}(x_2,0,x_4,1,x_3) - \mathbf{G}(x_2,0,x_4,0,x_3)
    \bigg\},
\end{align}
\begin{align} 
    F_R( x_1,x_2,x_3,x_4 ) = \frac 1 2 \frac{1}{x_1-x_2} 
    \bigg\{ 
        (x_1+x_3-1) &\Big[ \mathbf{G}(x_1,1,x_4,1,x_3) - \mathbf{G}(x_1,0,x_4,1,x_3) \Big] 
        \nonumber \\ \nonumber
        -(x_1+x_3) &\Big[ \mathbf{G}(x_1,1,x_4,0,x_3) - \mathbf{G}(x_1,0,x_4,0,x_3) \Big]
        \\ \nonumber
        -(x_2+x_3-1) &\Big[ \mathbf{G}(x_2,1,x_4,1,x_3) - \mathbf{G}(x_2,0,x_4,1,x_3) \Big]
        \\ \nonumber
        + (x_2+x_3) &\Big[ \mathbf{G}(x_2,1,x_4,0,x_3) - \mathbf{G}(x_2,0,x_4,0,x_3) \Big] 
        \\
        +\Big[ \mathbf{A}(1) - \mathbf{A}(0) \Big] &\Big[ \mathbf{I}(x_1,1,x_4) - \mathbf{I}(x_1,0,x_4) - \mathbf{I}(x_2,1,x_4) + \mathbf{I}(x_2,0,x_4) \Big]
    \bigg\}.
\end{align}
One can check that the loop integral decompositions are still
symmetric under the interchange of $x_1$ and $x_2$, as it was the
case with the original integral definitions in eq. \eqref{eq:Floop model5}.

\bibliographystyle{BibFiles/t1}
\bibliography{BibFiles/MyBibTexDatabase}

\begin{thebibliography}{10}
\providecommand{\url}[1]{\texttt{#1}}
\providecommand{\urlprefix}{URL }
\providecommand{\eprint}[2][]{\url{#2}}

\bibitem{Weinberg:1979sa}
S.~Weinberg, \emph{{Baryon and Lepton Nonconserving Processes}},
  \MYhref[journalLinks]{http://dx.doi.org/10.1103/PhysRevLett.43.1566}{Phys.
  Rev. Lett.
  }\MYhref[journalLinks]{http://dx.doi.org/10.1103/PhysRevLett.43.1566}{\textbf{43}
  (1979) 1566--1570}.

\bibitem{Minkowski:1977sc}
P.~Minkowski, \emph{{$\mu\rightarrow e\gamma$ at a rate of one out of $10^9$
  muon decays?}},
  \MYhref[journalLinks]{http://dx.doi.org/10.1016/0370-2693(77)90435-X}{Phys.Lett.
  }\MYhref[journalLinks]{http://dx.doi.org/10.1016/0370-2693(77)90435-X}{\textbf{B67}
  (1977) 421}.

\bibitem{Yanagida:1979as}
T.~Yanagida, \emph{{Horizontal symmetry and masses of neutrinos}}, Workshop on
  the baryon number of the Universe and unified theories, O. Sawada and A.
  Sugamoto, eds.  (1979) 95.

\bibitem{Mohapatra:1979ia}
R.~N. Mohapatra and G.~Senjanovic, \emph{{Neutrino mass and spontaneous parity
  violation}},
  \MYhref[journalLinks]{http://dx.doi.org/10.1103/PhysRevLett.44.912}{Phys.
  Rev. Lett.
  }\MYhref[journalLinks]{http://dx.doi.org/10.1103/PhysRevLett.44.912}{\textbf{44}
  (1980) 912}.

\bibitem{Schechter:1980gr}
J.~Schechter and J.~Valle, \emph{{Neutrino masses in $SU(2) \times U(1)$
  theories}},
  \MYhref[journalLinks]{http://dx.doi.org/10.1103/PhysRevD.22.2227}{Phys. Rev.
  }\MYhref[journalLinks]{http://dx.doi.org/10.1103/PhysRevD.22.2227}{\textbf{D22}
  (1980) 2227}.

\bibitem{Anamiati:2018cuq}
G.~Anamiati et~al., \emph{{High-dimensional neutrino masses}}  (2018),
  \MYhref[eprintLinks]{http://arxiv.org/abs/1806.07264}{{\ttfamily
  arXiv:1806.07264 [hep-ph]}}.

\bibitem{Zee:1980ai}
A.~Zee, \emph{{A theory of lepton number violation, neutrino Majorana mass, and
  oscillation}},
  \MYhref[journalLinks]{http://dx.doi.org/10.1016/0370-2693(80)90349-4,
  10.1016/0370-2693(80)90349-4}{Phys.Lett.
  }\MYhref[journalLinks]{http://dx.doi.org/10.1016/0370-2693(80)90349-4,
  10.1016/0370-2693(80)90349-4}{\textbf{B93} (1980) 389}.

\bibitem{Cai:2017jrq}
Y.~Cai et~al., \emph{{From the trees to the forest: a review of radiative
  neutrino mass models}},
  \MYhref[journalLinks]{http://dx.doi.org/10.3389/fphy.2017.00063}{Front.in
  Phys.
  }\MYhref[journalLinks]{http://dx.doi.org/10.3389/fphy.2017.00063}{\textbf{5}
  (2017) 63}, \MYhref[eprintLinks]{http://arxiv.org/abs/1706.08524}{{\ttfamily
  arXiv:1706.08524 [hep-ph]}}.

\bibitem{Farzan:2012ev}
Y.~Farzan, S.~Pascoli and M.~A. Schmidt, \emph{{Recipes and ingredients for
  neutrino mass at loop level}},
  \MYhref[journalLinks]{http://dx.doi.org/10.1007/JHEP03(2013)107}{JHEP
  }\MYhref[journalLinks]{http://dx.doi.org/10.1007/JHEP03(2013)107}{\textbf{03}
  (2013) 107}, \MYhref[eprintLinks]{http://arxiv.org/abs/1208.2732}{{\ttfamily
  arXiv:1208.2732 [hep-ph]}}.

\bibitem{Ma:1998dn}
E.~Ma, \emph{{Pathways to naturally small neutrino masses}},
  \MYhref[journalLinks]{http://dx.doi.org/10.1103/PhysRevLett.81.1171}{Phys.Rev.Lett.
  }\MYhref[journalLinks]{http://dx.doi.org/10.1103/PhysRevLett.81.1171}{\textbf{81}
  (1998) 1171--1174},
  \MYhref[eprintLinks]{http://arxiv.org/abs/hep-ph/9805219}{{\ttfamily
  arXiv:hep-ph/9805219 [hep-ph]}}.

\bibitem{Bonnet:2012kz}
F.~Bonnet, M.~Hirsch, T.~Ota and W.~Winter, \emph{{Systematic study of the
  $d=5$ Weinberg operator at one-loop order}},
  \MYhref[journalLinks]{http://dx.doi.org/10.1007/JHEP07(2012)153}{JHEP
  }\MYhref[journalLinks]{http://dx.doi.org/10.1007/JHEP07(2012)153}{\textbf{1207}
  (2012) 153}, \MYhref[eprintLinks]{http://arxiv.org/abs/1204.5862}{{\ttfamily
  arXiv:1204.5862 [hep-ph]}}.

\bibitem{Sierra:2014rxa}
D.~Aristizabal~Sierra, A.~Degee, L.~Dorame and M.~Hirsch, \emph{{Systematic
  classification of two-loop realizations of the Weinberg operator}},
  \MYhref[journalLinks]{http://dx.doi.org/10.1007/JHEP03(2015)040}{JHEP
  }\MYhref[journalLinks]{http://dx.doi.org/10.1007/JHEP03(2015)040}{\textbf{03}
  (2015) 040}, \MYhref[eprintLinks]{http://arxiv.org/abs/1411.7038}{{\ttfamily
  arXiv:1411.7038 [hep-ph]}}.

\bibitem{Bonnet:2009ej}
F.~Bonnet, D.~Hernandez, T.~Ota and W.~Winter, \emph{{Neutrino masses from
  higher than $d=5$ effective operators}},
  \MYhref[journalLinks]{http://dx.doi.org/10.1088/1126-6708/2009/10/076}{JHEP
  }\MYhref[journalLinks]{http://dx.doi.org/10.1088/1126-6708/2009/10/076}{\textbf{0910}
  (2009) 076}, \MYhref[eprintLinks]{http://arxiv.org/abs/0907.3143}{{\ttfamily
  arXiv:0907.3143 [hep-ph]}}.

\bibitem{Cepedello:2017eqf}
R.~Cepedello, M.~Hirsch and J.~C. Helo, \emph{{Loop neutrino masses from $d =
  7$ operator}},
  \MYhref[journalLinks]{http://dx.doi.org/10.1007/JHEP07(2017)079}{JHEP
  }\MYhref[journalLinks]{http://dx.doi.org/10.1007/JHEP07(2017)079}{\textbf{07}
  (2017) 079}, \MYhref[eprintLinks]{http://arxiv.org/abs/1705.01489}{{\ttfamily
  arXiv:1705.01489 [hep-ph]}}.

\bibitem{Babu:2009aq}
K.~S. Babu, S.~Nandi and Z.~Tavartkiladze, \emph{{New mechanism for neutrino
  mass generation and triply charged Higgs bosons at the LHC}},
  \MYhref[journalLinks]{http://dx.doi.org/10.1103/PhysRevD.80.071702}{Phys.
  Rev.
  }\MYhref[journalLinks]{http://dx.doi.org/10.1103/PhysRevD.80.071702}{\textbf{D80}
  (2009) 071702},
  \MYhref[eprintLinks]{http://arxiv.org/abs/0905.2710}{{\ttfamily
  arXiv:0905.2710 [hep-ph]}}.

\bibitem{Babu:2001ex}
K.~S. Babu and C.~N. Leung, \emph{{Classification of effective neutrino mass
  operators}},
  \MYhref[journalLinks]{http://dx.doi.org/10.1016/S0550-3213(01)00504-1}{Nucl.
  Phys.
  }\MYhref[journalLinks]{http://dx.doi.org/10.1016/S0550-3213(01)00504-1}{\textbf{B619}
  (2001) 667--689},
  \MYhref[eprintLinks]{http://arxiv.org/abs/hep-ph/0106054}{{\ttfamily
  arXiv:hep-ph/0106054 [hep-ph]}}.

\bibitem{deGouvea:2007qla}
A.~de~Gouvea and J.~Jenkins, \emph{{A Survey of Lepton Number Violation Via
  Effective Operators}},
  \MYhref[journalLinks]{http://dx.doi.org/10.1103/PhysRevD.77.013008}{Phys.
  Rev.
  }\MYhref[journalLinks]{http://dx.doi.org/10.1103/PhysRevD.77.013008}{\textbf{D77}
  (2008) 013008},
  \MYhref[eprintLinks]{http://arxiv.org/abs/0708.1344}{{\ttfamily
  arXiv:0708.1344 [hep-ph]}}.

\bibitem{Angel:2012ug}
P.~W. Angel, N.~L. Rodd and R.~R. Volkas, \emph{{Origin of neutrino masses at
  the LHC: $\Delta L = 2$ effective operators and their ultraviolet
  completions}},
  \MYhref[journalLinks]{http://dx.doi.org/10.1103/PhysRevD.87.073007}{Phys.
  Rev.
  }\MYhref[journalLinks]{http://dx.doi.org/10.1103/PhysRevD.87.073007}{\textbf{D87}
  (2013) 7 073007},
  \MYhref[eprintLinks]{http://arxiv.org/abs/1212.6111}{{\ttfamily
  arXiv:1212.6111 [hep-ph]}}.

\bibitem{Krauss:2002px}
L.~M. Krauss, S.~Nasri and M.~Trodden, \emph{{A model for neutrino masses and
  dark matter}},
  \MYhref[journalLinks]{http://dx.doi.org/10.1103/PhysRevD.67.085002}{Phys.
  Rev.
  }\MYhref[journalLinks]{http://dx.doi.org/10.1103/PhysRevD.67.085002}{\textbf{D67}
  (2003) 085002},
  \MYhref[eprintLinks]{http://arxiv.org/abs/hep-ph/0210389}{{\ttfamily
  arXiv:hep-ph/0210389 [hep-ph]}}.

\bibitem{Cheung:2004xm}
K.~Cheung and O.~Seto, \emph{{Phenomenology of TeV right-handed neutrino and
  the dark matter model}},
  \MYhref[journalLinks]{http://dx.doi.org/10.1103/PhysRevD.69.113009}{Phys.
  Rev.
  }\MYhref[journalLinks]{http://dx.doi.org/10.1103/PhysRevD.69.113009}{\textbf{D69}
  (2004) 113009},
  \MYhref[eprintLinks]{http://arxiv.org/abs/hep-ph/0403003}{{\ttfamily
  arXiv:hep-ph/0403003 [hep-ph]}}.

\bibitem{Ahriche:2014cda}
A.~Ahriche, C.-S. Chen, K.~L. McDonald and S.~Nasri, \emph{{Three-loop model of
  neutrino mass with dark matter}},
  \MYhref[journalLinks]{http://dx.doi.org/10.1103/PhysRevD.90.015024}{Phys.
  Rev.
  }\MYhref[journalLinks]{http://dx.doi.org/10.1103/PhysRevD.90.015024}{\textbf{D90}
  (2014) 015024},
  \MYhref[eprintLinks]{http://arxiv.org/abs/1404.2696}{{\ttfamily
  arXiv:1404.2696 [hep-ph]}}.

\bibitem{Ahriche:2014oda}
A.~Ahriche, K.~L. McDonald and S.~Nasri, \emph{{A Model of radiative neutrino
  mass: with or without dark matter}},
  \MYhref[journalLinks]{http://dx.doi.org/10.1007/JHEP10(2014)167}{JHEP
  }\MYhref[journalLinks]{http://dx.doi.org/10.1007/JHEP10(2014)167}{\textbf{10}
  (2014) 167}, \MYhref[eprintLinks]{http://arxiv.org/abs/1404.5917}{{\ttfamily
  arXiv:1404.5917 [hep-ph]}}.

\bibitem{Chen:2014ska}
C.-S. Chen, K.~L. McDonald and S.~Nasri, \emph{{A class of three-loop models
  with neutrino mass and dark matter}},
  \MYhref[journalLinks]{http://dx.doi.org/10.1016/j.physletb.2014.05.082}{Phys.
  Lett.
  }\MYhref[journalLinks]{http://dx.doi.org/10.1016/j.physletb.2014.05.082}{\textbf{B734}
  (2014) 388--393},
  \MYhref[eprintLinks]{http://arxiv.org/abs/1404.6033}{{\ttfamily
  arXiv:1404.6033 [hep-ph]}}.

\bibitem{Gu:2012tn}
P.-H. Gu, \emph{{From dark matter to neutrinoless double beta decay}}  (2012),
  \MYhref[eprintLinks]{http://arxiv.org/abs/1203.4165}{{\ttfamily
  arXiv:1203.4165 [hep-ph]}}.

\bibitem{Nomura:2016ezz}
T.~Nomura, H.~Okada and N.~Okada, \emph{{A Colored KNT neutrino model}},
  \MYhref[journalLinks]{http://dx.doi.org/10.1016/j.physletb.2016.09.038}{Phys.
  Lett.
  }\MYhref[journalLinks]{http://dx.doi.org/10.1016/j.physletb.2016.09.038}{\textbf{B762}
  (2016) 409--414},
  \MYhref[eprintLinks]{http://arxiv.org/abs/1608.02694}{{\ttfamily
  arXiv:1608.02694 [hep-ph]}}.

\bibitem{Cheung:2016frv}
K.~Cheung, T.~Nomura and H.~Okada, \emph{{Three-loop neutrino mass model with a
  colored triplet scalar}},
  \MYhref[journalLinks]{http://dx.doi.org/10.1103/PhysRevD.95.015026}{Phys.
  Rev.
  }\MYhref[journalLinks]{http://dx.doi.org/10.1103/PhysRevD.95.015026}{\textbf{D95}
  (2017) 1 015026},
  \MYhref[eprintLinks]{http://arxiv.org/abs/1610.04986}{{\ttfamily
  arXiv:1610.04986 [hep-ph]}}.

\bibitem{Okada:2016rav}
H.~Okada and K.~Yagyu, \emph{{Renormalizable model for neutrino mass, dark
  matter, muon $g-2$ and 750 GeV diphoton excess}},
  \MYhref[journalLinks]{http://dx.doi.org/10.1016/j.physletb.2016.03.040}{Phys.
  Lett.
  }\MYhref[journalLinks]{http://dx.doi.org/10.1016/j.physletb.2016.03.040}{\textbf{B756}
  (2016) 337--344},
  \MYhref[eprintLinks]{http://arxiv.org/abs/1601.05038}{{\ttfamily
  arXiv:1601.05038 [hep-ph]}}.

\bibitem{Ahriche:2015loa}
A.~Ahriche, K.~L. McDonald and S.~Nasri, \emph{{A radiative model for the weak
  scale and neutrino mass via dark matter}},
  \MYhref[journalLinks]{http://dx.doi.org/10.1007/JHEP02(2016)038}{JHEP
  }\MYhref[journalLinks]{http://dx.doi.org/10.1007/JHEP02(2016)038}{\textbf{02}
  (2016) 038}, \MYhref[eprintLinks]{http://arxiv.org/abs/1508.02607}{{\ttfamily
  arXiv:1508.02607 [hep-ph]}}.

\bibitem{Ahriche:2015taa}
A.~Ahriche, K.~L. McDonald and S.~Nasri, \emph{{Scalar sector phenomenology of
  three-loop radiative neutrino mMass models}},
  \MYhref[journalLinks]{http://dx.doi.org/10.1103/PhysRevD.92.095020}{Phys.
  Rev.
  }\MYhref[journalLinks]{http://dx.doi.org/10.1103/PhysRevD.92.095020}{\textbf{D92}
  (2015) 9 095020},
  \MYhref[eprintLinks]{http://arxiv.org/abs/1508.05881}{{\ttfamily
  arXiv:1508.05881 [hep-ph]}}.

\bibitem{Ahriche:2015lqa}
A.~Ahriche, K.~L. McDonald and S.~Nasri, \emph{{Three-loop neutrino mass models
  at colliders}}, in \emph{{Proceedings, 50th Rencontres de Moriond Electroweak
  Interactions and Unified Theories: La Thuile, Italy, March 14-21, 2015}}
  (2015)  pages 285--290,
  \MYhref[eprintLinks]{http://arxiv.org/abs/1505.04320}{{\ttfamily
  arXiv:1505.04320 [hep-ph]}},
  \urlprefix\url{http://inspirehep.net/record/1370679/files/arXiv:1505.04320.pdf}.

\bibitem{Ahriche:2014xra}
A.~Ahriche, S.~Nasri and R.~Soualah, \emph{{Radiative neutrino mass model at
  the $e^-e^+$ linear collider}},
  \MYhref[journalLinks]{http://dx.doi.org/10.1103/PhysRevD.89.095010}{Phys.
  Rev.
  }\MYhref[journalLinks]{http://dx.doi.org/10.1103/PhysRevD.89.095010}{\textbf{D89}
  (2014) 9 095010},
  \MYhref[eprintLinks]{http://arxiv.org/abs/1403.5694}{{\ttfamily
  arXiv:1403.5694 [hep-ph]}}.

\bibitem{Ahriche:2015wha}
A.~Ahriche, K.~L. McDonald, S.~Nasri and T.~Toma, \emph{{A model of neutrino
  mass and dark matter with an accidental symmetry}},
  \MYhref[journalLinks]{http://dx.doi.org/10.1016/j.physletb.2015.05.031}{Phys.
  Lett.
  }\MYhref[journalLinks]{http://dx.doi.org/10.1016/j.physletb.2015.05.031}{\textbf{B746}
  (2015) 430--435},
  \MYhref[eprintLinks]{http://arxiv.org/abs/1504.05755}{{\ttfamily
  arXiv:1504.05755 [hep-ph]}}.

\bibitem{Hati:2018fzc}
C.~Hati, G.~Kumar, J.~Orloff and A.~M. Teixeira, \emph{{Reconciling $B$-decay
  anomalies with neutrino masses, dark matter and constraints from flavour
  violation}}  (2018),
  \MYhref[eprintLinks]{http://arxiv.org/abs/1806.10146}{{\ttfamily
  arXiv:1806.10146 [hep-ph]}}.

\bibitem{Aoki:2008av}
M.~Aoki, S.~Kanemura and O.~Seto, \emph{{Neutrino mass, dark matter and baryon
  asymmetry via TeV-scale physics without fine-tuning}},
  \MYhref[journalLinks]{http://dx.doi.org/10.1103/PhysRevLett.102.051805}{Phys.
  Rev. Lett.
  }\MYhref[journalLinks]{http://dx.doi.org/10.1103/PhysRevLett.102.051805}{\textbf{102}
  (2009) 051805},
  \MYhref[eprintLinks]{http://arxiv.org/abs/0807.0361}{{\ttfamily
  arXiv:0807.0361 [hep-ph]}}.

\bibitem{Aoki:2009vf}
M.~Aoki, S.~Kanemura and O.~Seto, \emph{{A model of TeV scale physics for
  neutrino mass, dark matter and baryon asymmetry and its phenomenology}},
  \MYhref[journalLinks]{http://dx.doi.org/10.1103/PhysRevD.80.033007}{Phys.
  Rev.
  }\MYhref[journalLinks]{http://dx.doi.org/10.1103/PhysRevD.80.033007}{\textbf{D80}
  (2009) 033007},
  \MYhref[eprintLinks]{http://arxiv.org/abs/0904.3829}{{\ttfamily
  arXiv:0904.3829 [hep-ph]}}.

\bibitem{Aoki:2010aq}
M.~Aoki, S.~Kanemura and O.~Seto, \emph{{ILC phenomenology in a TeV scale
  radiative seesaw model for neutrino mass, dark matter and baryon asymmetry}},
  in \emph{{8th General Meeting of the ILC Physics Subgroup Tsukuba, Japan,
  January 21, 2009}} (2010)
  \MYhref[eprintLinks]{http://arxiv.org/abs/1008.2407}{{\ttfamily
  arXiv:1008.2407 [hep-ph]}},
  \urlprefix\url{http://inspirehep.net/record/865435/files/arXiv:1008.2407.pdf}.

\bibitem{Aoki:2011zg}
M.~Aoki, S.~Kanemura and K.~Yagyu, \emph{{Triviality and vacuum stability
  bounds in the three-loop neutrino mass model}},
  \MYhref[journalLinks]{http://dx.doi.org/10.1103/PhysRevD.83.075016}{Phys.
  Rev.
  }\MYhref[journalLinks]{http://dx.doi.org/10.1103/PhysRevD.83.075016}{\textbf{D83}
  (2011) 075016},
  \MYhref[eprintLinks]{http://arxiv.org/abs/1102.3412}{{\ttfamily
  arXiv:1102.3412 [hep-ph]}}.

\bibitem{Okada:2015hia}
H.~Okada and K.~Yagyu, \emph{{Three-loop neutrino mass model with doubly
  charged particles from isodoublets}},
  \MYhref[journalLinks]{http://dx.doi.org/10.1103/PhysRevD.93.013004}{Phys.
  Rev.
  }\MYhref[journalLinks]{http://dx.doi.org/10.1103/PhysRevD.93.013004}{\textbf{D93}
  (2016) 1 013004},
  \MYhref[eprintLinks]{http://arxiv.org/abs/1508.01046}{{\ttfamily
  arXiv:1508.01046 [hep-ph]}}.

\bibitem{Ko:2016sxg}
P.~Ko, T.~Nomura, H.~Okada and Y.~Orikasa, \emph{{Confronting a new three-loop
  seesaw model with the 750 GeV diphoton excess}},
  \MYhref[journalLinks]{http://dx.doi.org/10.1103/PhysRevD.94.013009}{Phys.
  Rev.
  }\MYhref[journalLinks]{http://dx.doi.org/10.1103/PhysRevD.94.013009}{\textbf{D94}
  (2016) 1 013009},
  \MYhref[eprintLinks]{http://arxiv.org/abs/1602.07214}{{\ttfamily
  arXiv:1602.07214 [hep-ph]}}.

\bibitem{Gu:2016xno}
P.-H. Gu, \emph{{High-scale leptogenesis with three-loop neutrino mass
  generation and dark matter}},
  \MYhref[journalLinks]{http://dx.doi.org/10.1007/JHEP04(2017)159}{JHEP
  }\MYhref[journalLinks]{http://dx.doi.org/10.1007/JHEP04(2017)159}{\textbf{04}
  (2017) 159}, \MYhref[eprintLinks]{http://arxiv.org/abs/1611.03256}{{\ttfamily
  arXiv:1611.03256 [hep-ph]}}.

\bibitem{Culjak:2015qja}
P.~Culjak, K.~Kumericki and I.~Picek, \emph{{Scotogenic R$\nu$MDM at three-loop
  level}},
  \MYhref[journalLinks]{http://dx.doi.org/10.1016/j.physletb.2015.03.062}{Phys.
  Lett.
  }\MYhref[journalLinks]{http://dx.doi.org/10.1016/j.physletb.2015.03.062}{\textbf{B744}
  (2015) 237--243},
  \MYhref[eprintLinks]{http://arxiv.org/abs/1502.07887}{{\ttfamily
  arXiv:1502.07887 [hep-ph]}}.

\bibitem{Hatanaka:2014tba}
H.~Hatanaka, K.~Nishiwaki, H.~Okada and Y.~Orikasa, \emph{{A three-loop
  neutrino model with global $U(1)$ Symmetry}},
  \MYhref[journalLinks]{http://dx.doi.org/10.1016/j.nuclphysb.2015.03.006}{Nucl.
  Phys.
  }\MYhref[journalLinks]{http://dx.doi.org/10.1016/j.nuclphysb.2015.03.006}{\textbf{B894}
  (2015) 268--283},
  \MYhref[eprintLinks]{http://arxiv.org/abs/1412.8664}{{\ttfamily
  arXiv:1412.8664 [hep-ph]}}.

\bibitem{Cheng:1980qt}
T.~P. Cheng and L.-F. Li, \emph{{Neutrino masses, mixings and oscillations in
  $SU(2) x U(1)$ models of electroweak interactions}},
  \MYhref[journalLinks]{http://dx.doi.org/10.1103/PhysRevD.22.2860}{Phys. Rev.
  }\MYhref[journalLinks]{http://dx.doi.org/10.1103/PhysRevD.22.2860}{\textbf{D22}
  (1980) 2860}.

\bibitem{Zee:1985id}
A.~Zee, \emph{{Quantum numbers of Majorana neutrino masses}},
  \MYhref[journalLinks]{http://dx.doi.org/10.1016/0550-3213(86)90475-X}{Nucl.
  Phys.
  }\MYhref[journalLinks]{http://dx.doi.org/10.1016/0550-3213(86)90475-X}{\textbf{B264}
  (1986) 99--110}.

\bibitem{Babu:1988ki}
K.~S. Babu, \emph{{Model of ``calculable'' Majorana neutrino masses}},
  \MYhref[journalLinks]{http://dx.doi.org/10.1016/0370-2693(88)91584-5}{Phys.
  Lett.
  }\MYhref[journalLinks]{http://dx.doi.org/10.1016/0370-2693(88)91584-5}{\textbf{B203}
  (1988) 132--136}.

\bibitem{Nishiwaki:2015iqa}
K.~Nishiwaki, H.~Okada and Y.~Orikasa, \emph{{Three loop neutrino model with
  isolated $k^{\pm\pm}$}},
  \MYhref[journalLinks]{http://dx.doi.org/10.1103/PhysRevD.92.093013}{Phys.
  Rev.
  }\MYhref[journalLinks]{http://dx.doi.org/10.1103/PhysRevD.92.093013}{\textbf{D92}
  (2015) 9 093013},
  \MYhref[eprintLinks]{http://arxiv.org/abs/1507.02412}{{\ttfamily
  arXiv:1507.02412 [hep-ph]}}.

\bibitem{Dutta:2018qei}
B.~Dutta, S.~Ghosh, I.~Gogoladze and T.~Li, \emph{{Three-loop neutrino masses
  via new massive gauge bosons from $E_6$ GUT}}  (2018),
  \MYhref[eprintLinks]{http://arxiv.org/abs/1805.01866}{{\ttfamily
  arXiv:1805.01866 [hep-ph]}}.

\bibitem{Cheung:2017efc}
K.~Cheung, T.~Nomura and H.~Okada, \emph{{A three-loop neutrino model with
  leptoquark tTriplet scalars}},
  \MYhref[journalLinks]{http://dx.doi.org/10.1016/j.physletb.2017.03.021}{Phys.
  Lett.
  }\MYhref[journalLinks]{http://dx.doi.org/10.1016/j.physletb.2017.03.021}{\textbf{B768}
  (2017) 359--364},
  \MYhref[eprintLinks]{http://arxiv.org/abs/1701.01080}{{\ttfamily
  arXiv:1701.01080 [hep-ph]}}.

\bibitem{Gustafsson:2012vj}
M.~Gustafsson, J.~M. No and M.~A. Rivera, \emph{{Predictive model for
  radiatively induced neutrino masses and mixings with dark matter}},
  \MYhref[journalLinks]{http://dx.doi.org/10.1103/PhysRevLett.110.211802,
  10.1103/PhysRevLett.112.259902}{Phys. Rev. Lett.
  }\MYhref[journalLinks]{http://dx.doi.org/10.1103/PhysRevLett.110.211802,
  10.1103/PhysRevLett.112.259902}{\textbf{110} (2013) 21 211802}, [Erratum:
  Phys. Rev. Lett.112,no.25,259902(2014)],
  \MYhref[eprintLinks]{http://arxiv.org/abs/1212.4806}{{\ttfamily
  arXiv:1212.4806 [hep-ph]}}.

\bibitem{Gustafsson:2014vpa}
M.~Gustafsson, J.~M. No and M.~A. Rivera, \emph{{Radiative neutrino mass
  generation linked to neutrino mixing and $0\nu\beta\beta$-decay
  predictions}},
  \MYhref[journalLinks]{http://dx.doi.org/10.1103/PhysRevD.90.013012}{Phys.
  Rev.
  }\MYhref[journalLinks]{http://dx.doi.org/10.1103/PhysRevD.90.013012}{\textbf{D90}
  (2014) 1 013012},
  \MYhref[eprintLinks]{http://arxiv.org/abs/1402.0515}{{\ttfamily
  arXiv:1402.0515 [hep-ph]}}.

\bibitem{extraData}
\emph{{Systematic classification of three-loop realizations of the Weinberg
  operator: Extra data}}, See supplementary material, available also at
  \url{http://renatofonseca.net/3loop-Weinberg-operator.php}, accessed:
  05/06/2019.

\bibitem{Read:1981}
R.~C. Read, \emph{A survey of graph generation techniques}, in K.~L. McAvaney
  (editor), \emph{Combinatorial Mathematics VIII}, Springer Berlin Heidelberg,
  Berlin, Heidelberg, ISBN 978-3-540-38792-3 (1981)  pages 77--89.

\bibitem{Martin:2016bgz}
S.~P. Martin and D.~G. Robertson, \emph{{Evaluation of the general 3-loop
  vacuum Feynman integral}},
  \MYhref[journalLinks]{http://dx.doi.org/10.1103/PhysRevD.95.016008}{Phys.
  Rev.
  }\MYhref[journalLinks]{http://dx.doi.org/10.1103/PhysRevD.95.016008}{\textbf{D95}
  (2017) 1 016008},
  \MYhref[eprintLinks]{http://arxiv.org/abs/1610.07720}{{\ttfamily
  arXiv:1610.07720 [hep-ph]}}.

\bibitem{Borowka:2017idc}
S.~Borowka et~al., \emph{{pySecDec: a toolbox for the numerical evaluation of
  multi-scale integrals}},
  \MYhref[journalLinks]{http://dx.doi.org/10.1016/j.cpc.2017.09.015}{Comput.
  Phys. Commun.
  }\MYhref[journalLinks]{http://dx.doi.org/10.1016/j.cpc.2017.09.015}{\textbf{222}
  (2018) 313--326},
  \MYhref[eprintLinks]{http://arxiv.org/abs/1703.09692}{{\ttfamily
  arXiv:1703.09692 [hep-ph]}}.

\bibitem{Langacker:2011db}
P.~Langacker and G.~Steigman, \emph{{Requiem for an FCHAMP? Fractionally
  CHArged, Massive Particle}},
  \MYhref[journalLinks]{http://dx.doi.org/10.1103/PhysRevD.84.065040}{Phys.
  Rev.
  }\MYhref[journalLinks]{http://dx.doi.org/10.1103/PhysRevD.84.065040}{\textbf{D84}
  (2011) 065040},
  \MYhref[eprintLinks]{http://arxiv.org/abs/1107.3131}{{\ttfamily
  arXiv:1107.3131 [hep-ph]}}.

\bibitem{McDonald:2003zj}
K.~L. McDonald and B.~H.~J. McKellar, \emph{{Evaluating the two loop diagram
  responsible for neutrino mass in Babu's model}}  (2003),
  \MYhref[eprintLinks]{http://arxiv.org/abs/hep-ph/0309270}{{\ttfamily
  arXiv:hep-ph/0309270 [hep-ph]}}.

\bibitem{vanderBij:1983bw}
J.~van~der Bij and M.~J.~G. Veltman, \emph{{Two loop large Higgs mass
  correction to the $\rho$-parameter}},
  \MYhref[journalLinks]{http://dx.doi.org/10.1016/0550-3213(84)90284-0}{Nucl.
  Phys.
  }\MYhref[journalLinks]{http://dx.doi.org/10.1016/0550-3213(84)90284-0}{\textbf{B231}
  (1984) 205--234}.

\bibitem{Freitas:2016zmy}
A.~Freitas, \emph{{Three-loop vacuum integrals with arbitrary masses}},
  \MYhref[journalLinks]{http://dx.doi.org/10.1007/JHEP11(2016)145}{JHEP
  }\MYhref[journalLinks]{http://dx.doi.org/10.1007/JHEP11(2016)145}{\textbf{11}
  (2016) 145}, \MYhref[eprintLinks]{http://arxiv.org/abs/1609.09159}{{\ttfamily
  arXiv:1609.09159 [hep-ph]}}.

\bibitem{Chetyrkin:1981qh}
K.~G. Chetyrkin and F.~V. Tkachov, \emph{{Integration by parts: the algorithm
  to calculate $\beta$-functions in 4 loops}},
  \MYhref[journalLinks]{http://dx.doi.org/10.1016/0550-3213(81)90199-1}{Nucl.
  Phys.
  }\MYhref[journalLinks]{http://dx.doi.org/10.1016/0550-3213(81)90199-1}{\textbf{B192}
  (1981) 159--204}.

\bibitem{Passarino:1978jh}
G.~Passarino and M.~J.~G. Veltman, \emph{{One loop corrections for $e^+ e^-$
  annihilation into $\mu^+ \mu^-$ in the Weinberg model}},
  \MYhref[journalLinks]{http://dx.doi.org/10.1016/0550-3213(79)90234-7}{Nucl.
  Phys.
  }\MYhref[journalLinks]{http://dx.doi.org/10.1016/0550-3213(79)90234-7}{\textbf{B160}
  (1979) 151--207}.

\end{thebibliography}

\clearpage
\newpage

\section*{Erratum}

\begin{figure}
    \begin{centering}
        \includegraphics[width=1\textwidth]{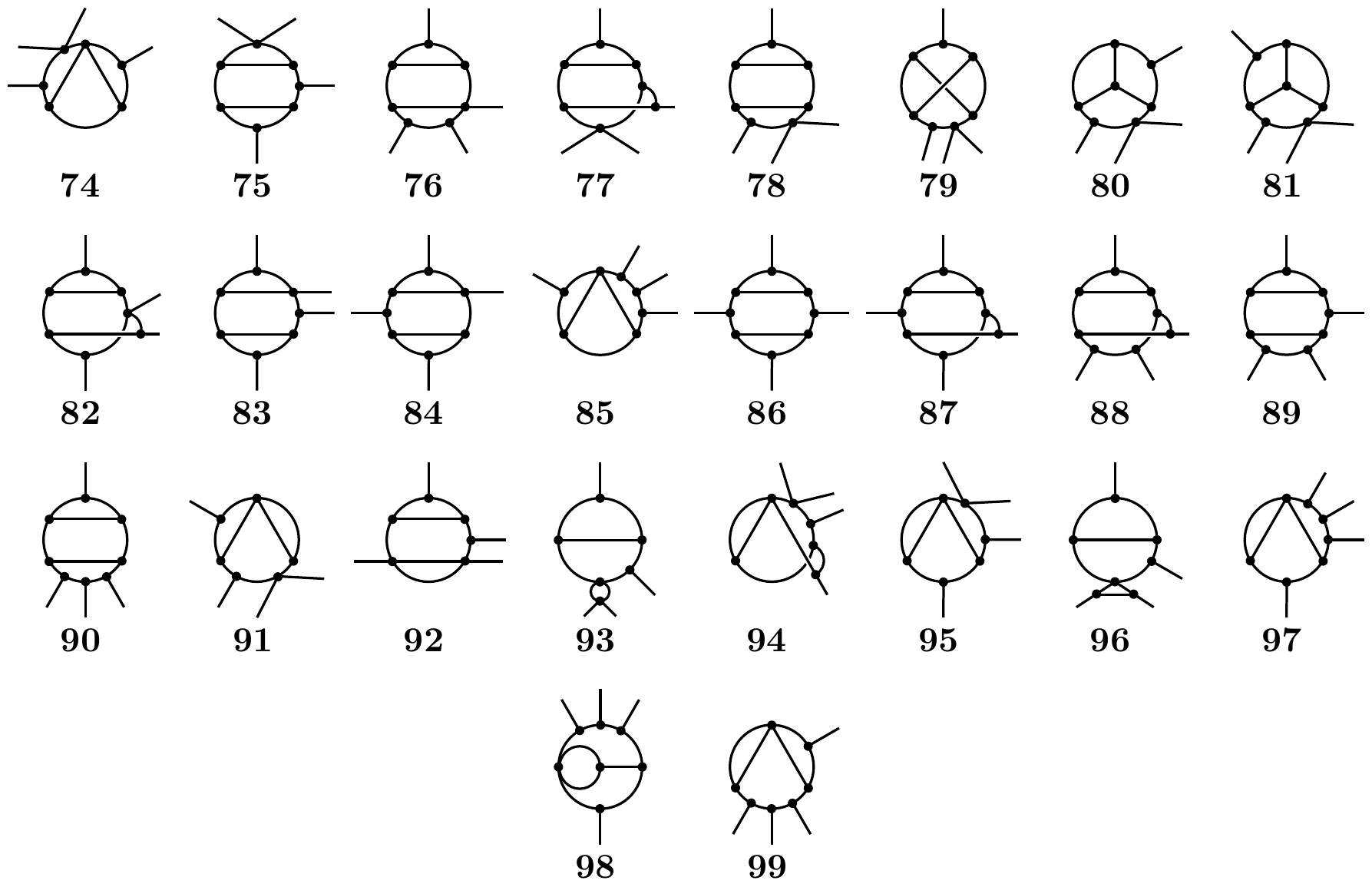}
        \par\end{centering} 
\protect\caption{\label{fig:topos}List of new topologies associated to
  \textit{special} genuine diagrams. We follow the numbering of the
  paper and this list should be understood as a continuation of
  figure~15 in the Appendix~A of the paper. See text for details.}
\end{figure}

The strategy used  in our original publication to classify the 3-loop realizations of the Weinberg operator
admits a loophole which was overlooked; it enlarges the set of \textit{genuine} 
topologies. We have found that by means of this loophole, the 26 topologies depicted in figure~\ref{fig:topos}
categorized as \textit{non-genuine} in our original publication
actually have to be classified as \textit{special genuine}.  Recall
that \textit{special genuine topologies} are those associated to
neutrino mass diagrams which under normal circumstances can be redrawn
with less loops, unless some particular quantum numbers are assigned
to some of the particles in the internal lines.  More specifically,
these special diagrams contain fermion-fermion-scalar,
$\left(\textrm{scalar}\right)^{3}$ and/or
$\left(\textrm{scalar}\right)^{4}$ effective interactions generated
through loops which cannot be compressed to a point, as is ordinarily
the case.  That is because these effective couplings involve
derivatives of the fields, making them non-renormalizable, so there
exist exceptional cases in which there is no corresponding tree level
realization of the effective vertex.
 
\begin{figure}
    \begin{centering}
        \includegraphics[scale=1.25]{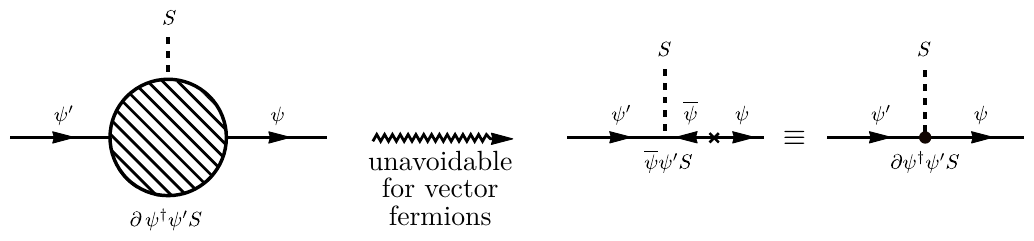}
\par\end{centering} 
    \protect\caption{\label{fig:explanation}Consider a loop induced
  coupling of the left-handed fermions $\psi$ and $\psi'$ to a scalar
  $S$ as indicated on the left. Given the chirality of the fermions,
  the effective coupling is $\partial \psi^\dagger \psi' S$. If $\psi$ has a vector-like partner
  $\overline{\psi}$ (which we may consider to be left-handed as well),
  then the tree-level coupling $\overline{\psi} \psi' S$ exists, and
  together with a mass insertion $\overline{\psi} \psi$ it can be used
  to generate the effective interaction $\partial
  \psi^\dagger \psi' S$ without loops (this is the leading order interaction; extra pairs of derivatives appear at higher order).  This argument fails if both
  $\psi$ and $\psi'$ are Standard Model fermions.}
\end{figure}

For the \textit{special genuine} topologies discussed in the paper,
the existence of derivatives can be traced to the antisymmetric nature
of some $SU(2)_L$ contractions, which makes some loop interactions
non-compressible for appropriate choices of the quantum numbers of the
diagram lines. However, when writing the paper we overlooked a second
possible reason why such derivative terms might be unavoidable:
fermion-fermion-scalar couplings may contain a derivative
due to the chirality of the standard model fermions.  In particular,
two left-handed Weyl fermions $\psi$ and $\psi'$ may interact with a scalar
$S$ through a $\partial\psi^\dagger \psi' S$ effective
coupling (the number of derivatives can be higher, as long as it is an odd number).\footnote{From a symmetry argument, one can see that the number of derivatives is odd, and therefore there is at least one of them. It goes as follows: the complexified algebra of the Lorentz group is the same as the complexified algebra of $SU(2)\times SU(2)$, so it's representations can be labeled by a pair of spins $(j_L,j_R)$. Left-handed fermions $\psi$ and their conjugates $\psi^\dagger$ transform as $(1/2,0)$ and $(0,1/2)$ (conjugation flips $j_L$ with $j_R$) respectively, while vector-like fields, such as derivatives, are bi-doublets $(1/2,1/2)$. Therefore, the bilinear $\psi \psi^\dagger$ transforms as a vector, and Lorentz invariance can only be obtained by adding to this fermion combination an odd number of derivatives.} However, if $\psi$ is a vector-like field, its left-handed
partner $\overline{\psi}$ has the same gauge quantum numbers as
$\psi^\dagger$ (the conjugate of $\psi$) and opposite chirality, hence there is no symmetry
forbidding the renormalizable interaction $\overline{\psi} \psi' S$,
with no derivatives.\footnote{The same argument follows interchanging
  $\psi$ with $\psi'$.} Using this coupling and a mass insertion
$\overline{\psi}\psi$ one can then always generate $\partial
\psi^\dagger \psi' S$ without loops, as depicted in
figure~\ref{fig:explanation}.  However, if neither $\psi$ nor $\psi'$
has a vector-like partner, this argument fails and the vertex
$\partial \psi^\dagger \psi' S$ may not be realizable at
tree-level. Thus, if $\psi$ and $\psi'$ are fixed to be Standard Model
fermions, the loop might not be compressible and our general 
arguments fail. Let us discuss this with one particular example.

\begin{figure}
    \begin{centering}
        \includegraphics[scale=1.25]{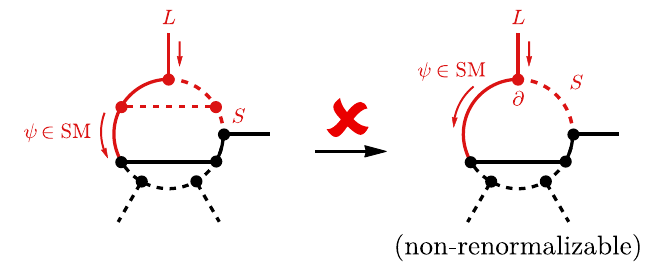}
\par\end{centering} 
    \protect\caption{\label{fig:example}Example of a diagram with topology
  89, containing a one-loop fermion-fermion-scalar effective
  interaction (in red). This loop is removable unless $\psi$ is a SM
  fermion.}
\end{figure}

Take for instance topology 89 in figure~\ref{fig:topos} and one of its
diagrams as an example (shown in figure~\ref{fig:example}).  This
diagram contains a one-loop realization of the vertex
$\partial L \psi^\dagger S$, with $L$ the SM lepton
doublet and $\psi$ and $S$ an arbitrary left-handed fermion and
scalar, respectively.  If $\psi$ is not a Standard Model fermion, it
is necessary to add its vector-like partner $\overline{\psi}$ to the
model, in order to generate a bare mass term $M
\overline{\psi}\psi$. From the argument in
figure~\ref{fig:explanation}, it is then possible to rewrite the
diagram with one less loop. On the other hand, it might not be
possible to do so if $\psi$ is a Standard Model fermion, in which
case the diagram is genuine.

Note that, since this loophole to our general argument exists only
for standard model fermions, the list of all possible genuine models
generated from the diagram in figure~\ref{fig:example} will be quite
constrained, due to the limited number of choices of $\psi \in
\{L,e^c,Q,u^c,d^c\}$.

\begin{figure}
	\begin{centering}
		\includegraphics[width=1\textwidth]{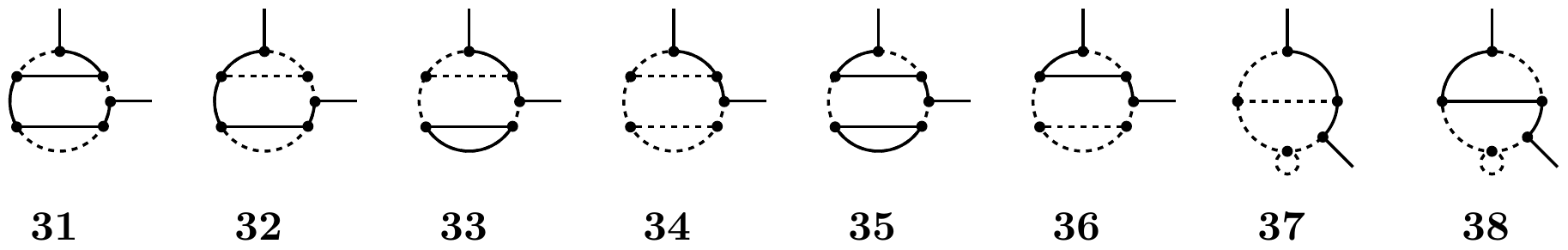}
		\par\end{centering} 
	\protect\caption{\label{fig:mass diag}List of new \textit{special}
  genuine diagrams in the mass basis, i.e. with the external Higgs
  lines removed. The numbering follows that of figure~8 of the paper
  and should be understood as a continuation of that figure.}
\end{figure}

In total, there are 125 genuine diagrams associated to the 26
topologies in figure~\ref{fig:topos}. They yield 8 new diagrams in the
mass basis, after electroweak symmetry breaking (see
figure~\ref{fig:mass diag}). The complete lists of topologies and
diagrams are given in \cite{extraData}.

\begin{figure}
    \begin{centering}
      \includegraphics[scale=0.8]{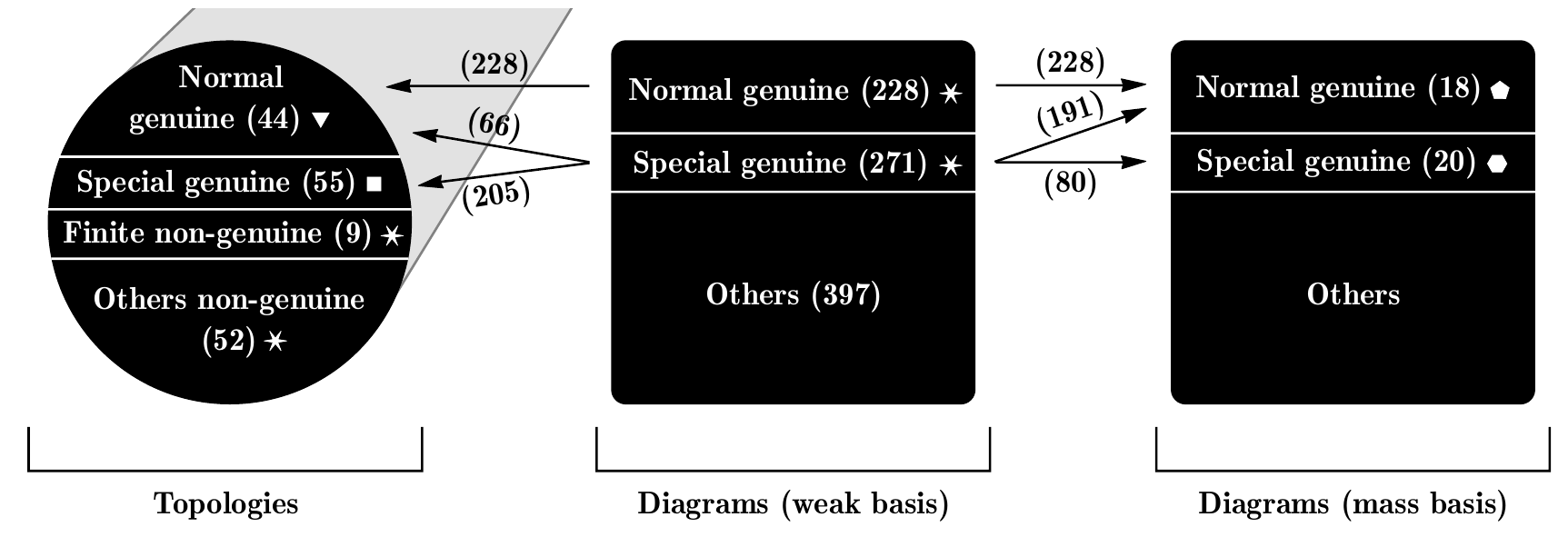}      
\par\end{centering} 
\protect\caption{\label{fig:fig9}Corrected counting for topologies
  and diagrams.}
\end{figure}

The newly re-classified topologies, see figure \ref{fig:topos}, and 
diagrams, figure \ref{fig:mass diag}, affect some numbers 
quoted in our original paper. In figure \ref{fig:fig9} we give the
updated numbers for each type of topology and each type of diagram.

\end{document}